\documentclass[%
aip,
jcp,%
amsmath,amssymb,
reprint,
groupedaddress,
longbibliography
]{revtex4-1}
\usepackage[latin1]{inputenc}

\usepackage{amsmath}%author-numerical,%
\usepackage{amsfonts}
\usepackage{amssymb,bm}
\usepackage{graphicx}
\usepackage{physics}
\usepackage{upgreek}
\usepackage[outercaption]{sidecap}   
\usepackage{color}
\usepackage[normalem]{ulem}
\mathchardef\mhyphen="2D
\usepackage{txfonts}

\DeclareMathAlphabet{\pazocal}{OMS}{zplm}{m}{n}

\usepackage{hyperref}

\begin{document}

\def\arctanh{\mathrm{arctanh}}
\def\expval#1#2{\bra{#2} {#1} \ket{#2}}
\def\mapright#1{\smash{\mathop{\longrightarrow}\limits^{_{_{\phantom{X}}}{#1}_{_{\phantom{X}}}}}}

\title{Perturbatively corrected ring-polymer instanton theory for accurate tunneling splittings} 
%\title{Efficient and accurate tunneling splittings from perturbatively corrected instanton theory}
%\title{High accuracy tunneling splittings: perturbatively corrected instanton theory} 
%\title{Anharmonic corrections to Instanton Theory}

\author{Joseph E. Lawrence}
\email{joseph.lawrence@phys.chem.ethz.ch}
\author{Jind\v{r}ich Du\v{s}ek}
\author{Jeremy O. Richardson}
\email{jeremy.richardson@phys.chem.ethz.ch}
\affiliation{Laboratory of Physical Chemistry, ETH Z\"urich, 8093 Z\"urich, Switzerland}

\date{January 2023}

\begin{abstract}
We introduce an approach for calculating perturbative corrections to the ring-polymer instanton approximation to tunneling splittings (RPI+PC), by computing higher-order terms in the asymptotic expansion in $\hbar$. The resulting method goes beyond 
%the harmonic approximation of
standard instanton theory by using information on the third and fourth derivatives of the potential along the tunneling path to include additional anharmonic effects.  This leads to significant improvements both in systems with low barriers and in systems with anharmonic modes.
%higher-order terms in the asymptotic expansion in $\hbar$
We demonstrate the applicability of RPI+PC to molecular systems by computing the tunneling splitting in full-dimensional malonaldehyde and a deuterated derivative. Comparing to both experiment and recent quantum-mechanical benchmark  results, we find that our perturbative correction reduces the error from $-$11\% to 2\% for hydrogen transfer and performs even better for the deuterated case. This makes our approach more accurate than previous calculations using diffusion Monte Carlo and path-integral molecular dynamics, while being more computationally efficient. 

% We find that errors are reduced by nearly an order of magnitude compared to the original instanton predictions, and obtain excellent agreement with both benchmark quantum mechanical results and experimental measurements.
\end{abstract}

\maketitle

\section{Introduction}
Tunneling splittings provide a unique window on the effect of quantum tunneling in chemical systems. The importance of quantum tunneling to processes in chemistry, from electron transfer in solution\cite{ChandlerET,Menzeleev2011ET,Fe2Fe3,Lawrence2020FeIIFeIII} to enzymatic reactions,\cite{HammesSchiffer2006EnzymeTunnel,Hu2014enzyme,Rommel2012enzyme} atmospheric\cite{Fang2016CH3CHOO,DoSTMI} and astrochemical\cite{Shannon2013tunnel} reactions to the structure and dynamics of water,\cite{Keutsch2001water,hexamerprism,WaterChapter,Ceriotti2016water} is now well established.  Unlike rate constants which are typically difficult to measure to high accuracy, tunneling splittings are spectroscopically measurable and hence they can often be determined to very high accuracy. This means that tunneling splittings present a stringent test to theory, requiring both an accurate description of interatomic interactions (potential energy surfaces (PES))\cite{StoneBook} as well as an accurate treatment of the resulting nuclear Hamiltonian.  Here we focus on the second of these two components, namely the computation of tunneling splittings given an accurate potential energy surface (whether fitted or calculated on the fly).

There exist a number of different approaches to calculating tunneling splittings in chemical systems. Conceptually, the simplest approach is to attempt to solve the nuclear Schr\"odinger equation on a grid. Unfortunately, due to the exponential scaling of the size of the Hilbert space these methods are limited to systems containing a handful of nuclei.\cite{Schroeder2011malonaldehyde,Hammer2011malonaldehyde,Carrington2017quantum,Stateresolved2022larsson} An alternative approach is provided by diffusion Monte Carlo (DMC),\cite{Suhm1991DMC,Gregory1995dmc} in which one propagates a stochastic representation of an initial trial wavefunction in imaginary time according to the Schr\"odinger equation to find the exact ground state, followed by a similar calculation to find the first excited state. Whilst it is in principle exact, DMC requires knowledge of the nodal plane in the first excited state, which must typically be approximated.\cite{Suhm1991DMC} Furthermore, it is a stochastic method and hence is prone to statistical errors. Together, these errors can be quite significant, especially for small splittings. Approaches based on the path-integral formulation of quantum mechanics\cite{Feynman} avoid the need to work directly with wavefunctions altogether by relating the tunneling splitting directly to the imaginary-time propagator. One can then use imaginary-time path-integral sampling methods such as path-integral molecular dynamics (PIMD)\cite{Parrinello1984Fcenter} and path-integral Monte Carlo (PIMC)\cite{Barker1979PI,Chandler+Wolynes1981} to obtain exact numerical calculation of the tunneling splitting.\cite{Ceperley1987exchange,Alexandrou1988tunnelling,Marchi1991tunnelling,Matyus2016tunnel1,Matyus2016tunnel2,Vaillant2018dimer,Vaillant2019water,Zhu2022trimer,Trenins2023tunnel} Both of these approaches make use of the so called ``classical isomorphism'': the equivalence of the discretized imaginary-time path integral to the classical statistical mechanics of a ring polymer made from many copies of the original system connected by harmonic springs. The difficulty with PIMD and PIMC approaches is that they require a large number of potential energy evaluations in order to reduce the associated statistical error and hence can become prohibitively expensive, especially when combined with high-level ab-initio potential energy evaluation. 

Due to the high cost of formally exact approaches it is often desirable to make use of approximate methods for estimating the tunneling splitting. The simplest of these approaches involve constructing simple one dimensional models, for example using the minimum-energy path, and then calculating the tunneling splitting using the Wentzel--Kramers--Brillouin (WKB) approximation.\cite{Landau+Lifshitz,Wales1993tunnel} Although computationally very cheap, these methods typically make large errors, as they neglect important multidimensional phenomena such as corner cutting. Recently it has been suggested that one can improve upon these simple models by using the ideas of vibrational perturbation theory (VPT)\cite{Mills1972VPT2} and semiclassical transition state theory (SCTST).\cite{Miller1990SCTST} In this approach, labeled semiclassical perturbation theory (SPT), one uses third and fourth derivatives of the potential at the minima and saddle point  to perturbatively include some of the effects missed by the simplest one-dimensional models.\cite{Burd2020tunnel} These approaches give reasonable results in certain regimes, however, just as SCTST breaks down for deep tunneling, SPT is unable to treat high barriers. An alternative approach, which maintains the computational efficiency of these simple models whilst generally providing much more accurate results is instanton theory.\cite{Uses_of_Instantons,Benderskii,Perspective,tunnel,InstReview}

Instanton theory, like the PIMD and PIMC approaches, is based on the path-integral formulation of quantum mechanics. However, rather than evaluating the path-integral numerically, instanton theory expands around the optimal tunneling path (the instanton), and then computes the resulting integrals analytically.\cite{Uses_of_Instantons,Benderskii}  Because of this, instanton theory only requires knowledge of the potential energy surface along the instanton path, making it far more computationally efficient than sampling approaches such as PIMD, PIMC or DMC which typically require global knowledge of the potential energy surface. Within the ring-polymer framework the instanton is simply a stationary point on the extended classical potential of the ring polymer, and hence can be found using standard optimization algorithms.\cite{Perspective,tunnel,InstReview} Ring-polymer instanton theory (RPI) is therefore more comparable in cost to simple one-dimensional theories. However, unlike other simple theories which rely on a pre-defined path, instanton theory automatically locates the optimal tunneling path in full-dimensionality along which the action (rather than the energy) is minimal. The instanton path thus gives both quantitative and qualitative information on the tunneling process, identifying the atoms involved in the tunneling, predicting isotopic effects, and giving insights into the importance of phenomena such as corner cutting.
%treating all degrees of freedom on an equal footing and hence giving insights into the importance of phenomena such as ``corner cutting''.
The balance of computational efficiency and accuracy provided by instanton theory mean it has been applied to a wide range of molecules and clusters, \cite{Milnikov2003,water,octamer,hexamerprism,Cvitas2016instanton,i-wat2,pentamer,formic,chiral,tropolone,Erakovic2020instanton}
and has even been generalized to treat asymmetric systems. \cite{asymtunnel,Erakovic2022instanton}

% as instanton theory does not require the input of a pre-defined path it is able to treat all degrees of freedom on an equal footing and hence captures important corner cutting effects. The instanton path thus allows one to visualize the tunneling mechanism in terms of a simple classical trajectory, giving qualitative as well as ... 

Instanton theory is, however, still an approximate method. Nevertheless, as it is based on rigorous asymptotic approximation to the path-integral expression, it is possible to go beyond the leading-order approximation to compute higher-order perturbative corrections. The development of a perturbatively corrected ring-polymer instanton theory (RPI+PC) that is applicable to molecular systems is the focus of the present paper. Performing such perturbative corrections is standard in many areas of theoretical physics (quantum field theory, string theory etc.), and is explored, for example in the case of the one-dimensional quartic double well in Ref.~\citenum{Kleinert}. However, so far these techniques have not been used within the ring-polymer instanton framework and hence have not to date been applied to compute anharmonic corrections to tunneling splittings in realistic multidimensional chemical systems. As the resulting theory involves perturbative corrections in terms of third and fourth derivatives of the potential, it is natural to compare it to other perturbative theories used in chemistry such as VPT2\cite{Miller1990SCTST} (and CASPT2\cite{Finley1998CASPT2}). As we shall see however the resulting theory is only superficially similar to these approaches. In particular, as the present theory is based on an asymptotic expansion of the path integral and not on a perturbative expansion of the energy eigenvalues it does not suffer from the difficulties associated with degeneracy seen in Rayleigh--Schr\"odinger perturbation theory.
Also note that our perturbative approach \emph{improves upon} standard instanton theory, in contrast to previous perturbative \emph{approximations to} instanton theory.\cite{Benderskii1997instanton,Smedarchina2012rainbow}

The paper is organized as follows. Section \ref{Exact_theory} summarizes how the tunneling splitting for molecular systems can be represented exactly in terms of path integrals. Section \ref{Instanton_theory_section} introduces the instanton approximation and then uses standard results from asymptotic analysis to derive perturbative corrections to the usual ring-polymer instanton theory expression for tunneling splittings. Section \ref{Results} illustrates the resulting perturbatively corrected instanton theory (RPI+PC) on two model problems, the one-dimensional quartic double well and a two-dimensional model with anharmonic vibrational modes, before demonstrating the applicability of the method with application to a full-dimensional description of malonaldehyde using the state-of-the-art PES of Ref.\citenum{Mizukami2014malonaldehyde}, which was fitted to CCSD(T)(F12*) data. Section \ref{Conclusion} concludes.

\section{Exact Quantum-Mechanical Theory}\label{Exact_theory}

Consider an $f$-dimensional system with a Hamiltonian
\begin{equation}
    \hat{H} = \sum_{\zeta=1}^{f} \frac{\hat{p}_\zeta^2}{2m_\zeta} + V(\hat{\bm{q}})
\end{equation}
for which the potential energy, $V(\bm{q})$, has two minima $\bm{q}_a$ and $\bm{q}_b$ that are related by symmetry, and where for simplicity we define the zero of energy such that $V(\bm{q}_a)=V(\bm{q}_b)=0$. Note that one can generalize to more than two symmetrically related minima, e.g., by building on the ideas of Refs.~\citenum{Matyus2016tunnel2} and \citenum{water}, but we leave the details of this for future work. %the generalization to systems with more than two symmetrically related minima can be achieved 

\subsection{Systems without rotational symmetry}
Before considering molecular systems which possess rotational and translational symmetry, we begin by considering systems for which the two minima $\bm{q}_a$ and $\bm{q}_b$ are related by a symmetry operation that is fixed in configuration space.  We can then group the eigenstates of the Hamiltonian into symmetric ($+$) and antisymmetric ($-$) states such that
\begin{equation}
    \hat{H} \ket{\psi_{k,\pm}} = E_{k,\pm}\ket{\psi_{k,\pm}}
\end{equation}
and 
\begin{equation}
    \braket{\bm{q}_a}{\psi_{k,\pm}} = \pm\braket{\bm{q}_b}{\psi_{k,\pm}}.
\end{equation}

\begin{figure}
    \centering
    \includegraphics{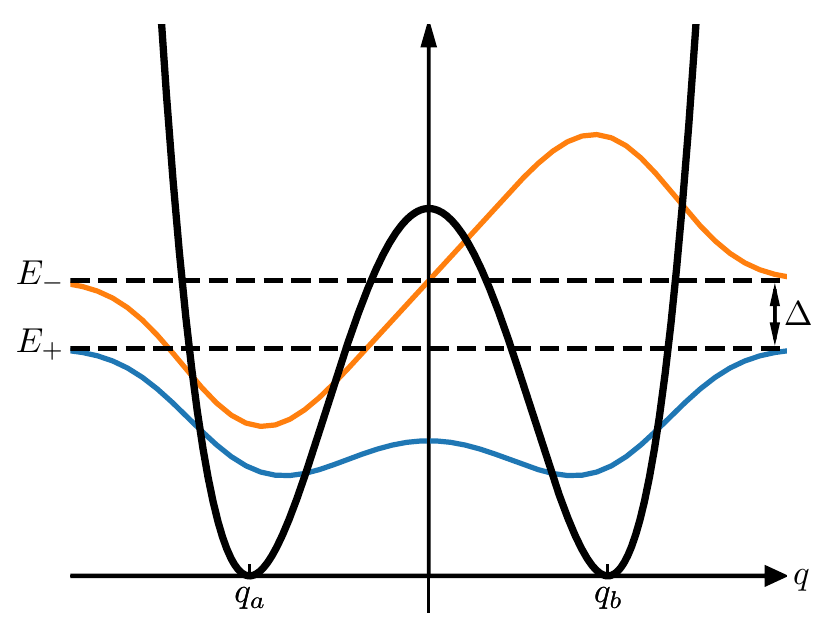}
    \caption{An illustration of a double-well system in one dimension, with the locations of the minima $q_a$ and $q_b$ labeled. The wavefunctions, $\psi_+$ and $\psi_-$, are shown at their corresponding energies, $E_+$ and $E_-$, and the magnitude of the tunneling splitting, $\Delta$, is also indicated with a double-headed arrow.}
    \label{fig:doublewell}
\end{figure}

We would like to calculate the energy splitting between the lowest symmetric and antisymmetric states $\Delta = E_{0,-}-E_{0,+}$. However, as the cost of explicitly constructing and diagonalizing the Hamiltonian can be very high, we would like to derive an expression which can be efficiently evaluated using a path-integral formalism. Hence we would like to obtain an expression for the splitting that depends on an imaginary-time propagator (or equivalently a density-matrix element with temperature $T=\hbar/k_B\tau$). This can be achieved (following Ref.~\citenum{Uses_of_Instantons} and Ref.~\citenum{Kleinert})%, and Ref.~\citenum{Ceperley1987exchange}
 by considering an imaginary-time propagator between the two minima 
\begin{equation}
\begin{aligned}
    \bra{\bm{q}_b} e^{-\tau\hat{H}/\hbar} \ket{\bm{q}_a} %&= \sum_{k,\sigma=\pm} \bra{\bm{q}_b} e^{-\tau\hat{H}/\hbar} \ket{\psi_{k,\sigma}} \braket{\psi_{k,\sigma}}{\bm{q}_a}\\
    &= \sum_{k,\sigma=\pm} \braket{\bm{q}_b}{\psi_{k,\sigma}} \braket{\psi_{k,\sigma}}{\bm{q}_a} e^{-\tau E_{k,\sigma}/\hbar}\\
    %&=\sum_{k,\sigma=\pm} \sigma\left|\braket{\bm{q}_a}{\psi_{k,\sigma}}\right|^2  e^{-\tau E_{k,\sigma}/\hbar}
    &=\sum_{k,\sigma=\pm} \sigma\left|\psi_{k,\sigma}(\bm{q}_a)\right|^2  e^{-\tau E_{k,\sigma}/\hbar}.
    \end{aligned}
\end{equation}
Then we take the limit that $\tau$ becomes large such that only the lowest states contribute significantly
% \begin{equation}
% \begin{aligned}
%     \bra{\bm{q}_a} e^{-\tau\hat{H}/\hbar} \ket{\bm{q}_b} &\simeq \left|\braket{\bm{q}_a}{\psi_{0,+}}\right|^2 e^{-\tau E_{0,+}/\hbar} -\left|\braket{\bm{q}_a}{\psi_{0,-}}\right|^2 e^{-\tau E_{0,-}/\hbar}.
%     \end{aligned}
% \end{equation}
\begin{equation}
\begin{aligned}
    \bra{\bm{q}_b} e^{-\tau\hat{H}/\hbar} \ket{\bm{q}_a} &= \left|\psi_{+}(\bm{q}_a)\right|^2 e^{-\tau E_{+}/\hbar} -\left|\psi_{-}(\bm{q}_a)\right|^2 e^{-\tau E_{-}/\hbar} ,
    \end{aligned}
\end{equation}
where we have dropped the subscript $0$ for notational simplicity.
Noting that the probability densities are similar but not identical we can rewrite them as $\left|\psi_{\pm}(\bm{q}_a)\right|^2=\rho e^{\pm\gamma}$. Also rewriting the energy eigenvalues as $E_{\pm}=E\mp\Delta/2$ we obtain
\begin{equation}
    \begin{aligned}
    \bra{\bm{q}_b} e^{-\tau\hat{H}/\hbar} \ket{\bm{q}_a} &= \rho  e^{-\tau E/\hbar} \left(e^{\tau\Delta/2\hbar+\gamma} -  e^{-\tau \Delta/2\hbar-\gamma}\right).
    \end{aligned}
\end{equation}
% \begin{equation}
%     \begin{aligned}
%     \bra{\bm{q}_a} e^{-\tau\hat{H}/\hbar} \ket{\bm{q}_b} &= \rho  e^{-\tau E_{+}/\hbar+\gamma} -\rho  e^{-\tau E_{-}/\hbar-\gamma}.
%     \end{aligned}
% \end{equation}
To arrive at an expression for the splitting $\Delta$ we then divide by the imaginary-time propagator which starts and ends in the same well to give
\begin{equation}
\begin{aligned}
    \frac{\bra{\bm{q}_b} e^{-\tau\hat{H}/\hbar} \ket{\bm{q}_a}}{\bra{\bm{q}_a} e^{-\tau\hat{H}/\hbar} \ket{\bm{q}_a}} 
%&=\frac{\rho  e^{-\tau E_{+}/\hbar+\gamma} -\rho  e^{-\tau E_{-}/\hbar-\gamma}}{\rho  e^{-\tau E_{+}/\hbar+\gamma} +\rho  e^{-\tau E_{-}/\hbar-\gamma}}\\
    &=\tanh\left(\frac{\tau\Delta}{2\hbar} + \gamma\right),
    \end{aligned}
\end{equation}
which is then rearranged to give an explicit (and so far exact) expression for the ground-state tunneling splitting
\begin{equation}
    \Delta = \lim_{\tau\to\infty}\frac{2\hbar}{\tau}\arctanh\left(\frac{\bra{\bm{q}_b} e^{-\tau\hat{H}/\hbar} \ket{\bm{q}_a}}{\bra{\bm{q}_a} e^{-\tau\hat{H}/\hbar} \ket{\bm{q}_a}}\right).
\end{equation}
The $\gamma$ dependence vanishes because it is a constant (independent of $\tau$) and hence $\gamma/\tau\to0$ as $\tau\to\infty$. Note that we have chosen this starting point for the instanton derivation, rather than the approach based on partition functions used in Refs.~\citenum{tunnel} and \citenum{Benderskii}, as it makes the derivation of the higher-order corrections to instanton theory more transparent. 

%[How, in principle, would you extend this to systems of multiple wells?]

%[derivation is similar to the one used by Coleman and  Kleinert - but differs from Benderskii and mine in that we don't do Q2 here]

\subsection{Systems with rotational symmetry}
For tunneling splittings of molecules in the gas phase one must additionally account for the rotational and translational symmetry of the Hamiltonian. The translational symmetry is trivial to deal with since the center-of-mass motion rigorously separates from the other degrees of freedom. However, for the rotational degrees of freedom we will need to formally project onto the appropriate rotational state,\cite{Matyus2016tunnel2,Vaillant2018instanton} which in the following will be taken to be the ground rotational state, $J=0$. Treating rotationally excited states is left for future work. The corresponding projection operator can be written as
\begin{equation}
    \hat{P}_{J=0} = \frac{1}{8\pi^2} \int \mathrm{d}\bm{q} \int \mathrm{d} \bm{\Omega} \ketbra{\bm{q}}{R_{\bm{\Omega}} (\bm{q})}
\end{equation}
where the function $R_{\bm{\Omega}}(\bm{q})$ returns $\bm{q}$ rotated by the Euler angles $\bm{\Omega}=(\chi,\theta,\phi)$ about its center of mass.

The derivation of the tunneling splitting then follows very similar steps to the case without rotational states. One again begins by defining the geometries of the two minima $\bm{q}_a$ and $\bm{q}_b$, both having their center of mass in the same location and a fixed relative rotational orientation---the arbitrary choice of the rotational orientation is unimportant due to the integration over $\bm{\Omega}$ in the projection onto $J=0$. The tunneling splitting can then be written as
\begin{equation}
\begin{aligned}
    \Delta &=   \lim_{\tau\to\infty}\frac{2\hbar}{\tau}\arctanh\left(\frac{\bra{\bm{q}_b}\hat{P}_{J=0} e^{-\tau\hat{H}/\hbar} \ket{\bm{q}_a}}{\bra{\bm{q}_a}\hat{P}_{J=0} e^{-\tau\hat{H}/\hbar} \ket{\bm{q}_a}}\right)\\
&=\lim_{\tau\to\infty}\frac{2\hbar}{\tau}\arctanh\left(\frac{\int \mathrm{d}\bm{\Omega}\bra{R_{\bm{\Omega}}(\bm{q}_b)} e^{-\tau\hat{H}/\hbar} \ket{\bm{q}_a}}{\int \mathrm{d}\bm{\Omega}\bra{R_{\bm{\Omega}}(\bm{q}_a)} e^{-\tau\hat{H}/\hbar} \ket{\bm{q}_a}}\right).
    \end{aligned}
\end{equation}

The integral over Euler angles is not always the most convenient expression for the tunneling splitting. One approach, employed in Ref.~\citenum{Vaillant2018instanton}, is to compute instantons for many values of $\bm{\Omega}$ and integrate by quadrature. Alternatively, (as is shown in Appendix \ref{Appendix_Euler_angles}) the explicit integration over Euler angles can be replaced with an integral over Cartesian positions to give 
\begin{equation}
    \Delta = \lim_{\tau\to\infty}\frac{2\hbar}{\tau}\arctanh\left(\frac{\int_b \mathrm{d}\bm{q}'\mel{\bm{q}'}{  e^{-\tau\hat{H}/\hbar} }{\bm{q}_a}}{\int_a\mathrm{d}\bm{q}'\mel{\bm{q}'}{   e^{-\tau\hat{H}/\hbar} }{\bm{q}_a}}\right) \label{Delta_J=0_q_version}
\end{equation}
in which $\int_a \mathrm{d}\bm{q}'=\int \mathrm{d}\bm{q}'\theta(-s({\bm{q}'}))$ and $\int_b \mathrm{d}\bm{q}'=\int \mathrm{d}\bm{q}'\theta(s({\bm{q}'}))$ are integrals restricted to the region around $\bm{q}_a$ and $\bm{q}_b$ respectively, with $\theta(x)$ being the Heaviside step function and $s(\bm{q})=0$ a dividing surface that separates the regions corresponding to the two minima. The absence of explicit integration over Euler angles makes this expression practically simpler to implement, which is why we shall use it in the remainder of this work. Note that, unlike the nodal surface of DMC, the exact form of $s(\bm{q})$ can be chosen arbitrarily and does not affect the final result.

\subsection{Path-integral expression}

The advantage of having written the tunneling splitting in terms of an expression involving imaginary-time propagators is that one can then recast these propagators in terms of a path integral. Performing a standard path-integral discretization we can write
\begin{equation}
    \int_c \mathrm{d}\bm{q}'\bra{\bm{q}'}  e^{-\tau\hat{H}/\hbar} \ket{\bm{q}_a}=\lim_{N\to\infty}A_N\int_c \mathrm{d}\mathbf{q} \, e^{-S(\mathbf{q})/\hbar} \label{Path-Integral}
\end{equation}
where $c=a$ or $b$ corresponds to the denominator or numerator of Eq.~\eqref{Delta_J=0_q_version}, and  
$\mathbf{q}=(\bm{q}_0=\bm{q}_a,\bm{q}_1,\dots,\bm{q}_{N-1},\bm{q}_{N}=\bm{q}')$ is a vector of the configurations of the system at each point along the path.
Note that the integral is performed only over $\bm{q}_1$ to $\bm{q}_N$ as $\bm{q}_0$ is fixed at $\bm{q}_a$, i.e.~$\int_c\mathrm{d}\mathbf{q}=\int_c\mathrm{d}\bm{q}_N\int\mathrm{d}\bm{q}_{N-1}\int\mathrm{d}\bm{q}_{N-2}\dots\int\mathrm{d}\bm{q}_{1}$. 
% Note we will make use of discrete path-integral notation throughout, as this is how the expressions are actually implemented.
As we are considering an imaginary-time propagator, the action, $S(\mathbf{q})$, is the discretized Euclidean action defined as
\begin{equation}
  S(\mathbf{q}) = \sum_{n=0}^{N-1}\left( \sum_{\zeta=1}^{f} m_\zeta\frac{(q_{\zeta,n+1}-q_{\zeta,n})^2}{2\delta\tau}+\delta\tau\frac{V(\bm{q}_{n+1})+V(\bm{q}_{n})}{2}\right)
\end{equation}
with $\delta\tau=\tau/N$. The prefactor $A_N$ is given by
\begin{equation}
    A_N =\prod_{\zeta=1}^{f}\left( \frac{ m_\zeta}{ 2\pi\hbar \delta \tau} \right)^{N/2}.
\end{equation}

Combining Eq.~\eqref{Delta_J=0_q_version} with Eq.~\eqref{Path-Integral} results in an exact expression for the tunneling splitting $\Delta$. Note that this is a subtly different expression to the ones used most commonly in the literature, as it involves a path integral in which one end is fixed and the other is free---as explained above this ensures the rotations are treated correctly. We could then evaluate this expression using path-integral molecular dynamics (PIMD) or path-integral Monte Carlo (PIMC) similarly to Refs.~\citenum{Ceperley1987exchange,Alexandrou1988tunnelling,Marchi1991tunnelling,Matyus2016tunnel1,Matyus2016tunnel2,Vaillant2018dimer,Vaillant2019water,Zhu2022trimer,Trenins2023tunnel}. Whilst this would be
 formally exact, PIMD and PIMC are statistical and hence require many samples to converge to the correct result. This means that to obtain accurate tunneling splittings via PIMD or PIMC a very large number of potential evaluations with high accuracy electronic structure methods are required. With machine-learning techniques reducing the cost of potential evaluations this is certainly not an impossible feat. However, there exists an alternative approach which is not only much more cost effective but also provides deeper mechanistic insight: instanton theory.

\section{Instanton theory and Asymptotic analysis}\label{Instanton_theory_section}
Instanton theory is part of a general class of rigorous semiclassical ($\hbar\to0$) methods. These methods are based on the techniques of asymptotic analysis,\cite{BenderBook} which is a powerful tool for obtaining numerically accurate approximations in terms of an expansion in a small parameter. By taking $\hbar$ as the small asymptotic parameter one obtains asymptotic approximations to path integrals in terms of classical trajectories, giving rise to the term semiclassical approximation. Whilst the resulting asymptotic expansion contains an infinite number of terms, the power of asymptotics is that one can typically obtain accurate results from just the first few terms in the series. Usually, in instanton theory one just takes the first term in the series. However, it is possible to increase the accuracy of the instanton approach by including higher-order terms in the series as perturbative corrections and this is what we will explore here. For readers unfamiliar with asymptotic analysis, a short introduction to the techniques used in the present work is given in Appendix \ref{Appendix:Intro_to_asymptotics}. Although an understanding of asymptotics is needed to obtain a full understanding of the mathematical justification of the method,  the following sections are written to be comprehensible for readers without a background in asymptotics.
% interested simply in the practical implementation of the method. 

\subsection{Semiclassical Approximation}

%Asymptotic analysis allows for the generalization of basic perturbative approaches to systems where it is not possible to obtain a formally convergent Taylor series expansion in the perturbation parameter. Instead one obtains what is known as an asymptotic series (of which Taylor series are a special case) the formal properties of which will be discussed later. Importantly asymptotic series can be used to obtain a very accurate approximation to the function of interest. 

Before we discuss perturbatively corrected instanton theory, we first give an overview of the standard (lowest order) instanton approach. The central idea behind instanton theory is to approximate the path integral using asymptotic techniques, e.g.~Laplace's method (explained in Appendices \ref{Appendix:One_dim_asymptotics} and \ref{Multidimensional_Asymptotics}). Practically this involves expanding the action up to second order about its minimum, $\tilde{\mathbf{q}}$ (where $\nabla S(\tilde{\bf{q}})=\bm{0}$ and $\tilde{\mathbf{q}}$ is restricted to be within the integration domain), 
\begin{equation}
    S(\mathbf{q}) = S(\tilde{\mathbf{q}}) + \frac{1}{2} (\mathbf{q}-\tilde{\mathbf{q}})^T S^{(2)}(\tilde{\mathbf{q}}) (\mathbf{q}-\tilde{\mathbf{q}})  + \cdots,
\end{equation}
where $S^{(2)}(\tilde{\mathbf{q}})$ denotes the matrix of second derivatives of $S$ evaluated at $\tilde{\mathbf{q}}$,
 and then performing the resulting Gaussian integrals analytically. Algebraically this corresponds to
 \begin{equation}
\begin{aligned}
    \int \mathrm{d}\mathbf{q} \, e^{-S(\mathbf{q})/\hbar} &\sim \int \mathrm{d}\mathbf{q} \, e^{-S(\tilde{\mathbf{q}})/\hbar - \frac{1}{2} (\mathbf{q}-\tilde{\mathbf{q}})^T S^{(2)}(\tilde{\mathbf{q}}) (\mathbf{q}-\tilde{\mathbf{q}})/\hbar} \\
    &= e^{-S(\tilde{\mathbf{q}})/\hbar}\det\left(\frac{S^{(2)}(\tilde{\mathbf{q}})}{2\pi\hbar}\right)^{-1/2}\\
    &=e^{-S(\tilde{\mathbf{q}})/\hbar}\det\left(\frac{\mathbf{D}}{2\pi\hbar}\right)^{-1/2}. \label{leading_order_inst}
    \end{aligned}
\end{equation}
In the final line we have replaced the second-derivative matrix with the diagonal matrix of its  eigenvalues $\mathbf{D}$. Note for later sections it is also useful to define the matrix of eigenvectors, $\mathbf{C}$, such that 
\begin{equation}
    \mathbf{C}^{T} S^{(2)}(\tilde{\mathbf{q}}) \mathbf{C} = \mathbf{D}.
\end{equation}

The resulting theory comes with a simple physical picture, as the minimum configuration, $\tilde{\bf{q}}$ is equivalent to a discretized classical trajectory. This can be seen by noting that the condition $\nabla S(\tilde{\bf{q}})=0$ can be rearranged to give
\begin{equation}
    \tilde{q}_{\zeta,n+1} = \tilde{q}_{\zeta,n} + \frac{\delta\tau}{m_\zeta}\frac{\tilde{q}_{\zeta,n}-\tilde{q}_{\zeta,n-1}}{\delta\tau} + \frac{(\delta\tau)^2}{m_\zeta}\frac{\partial V}{\partial \tilde{q}_{\zeta,n}},
\end{equation}
which is equivalent to the Verlet integrator for Newton's equations of motion in imaginary time, with the identification of $\bm{q}_n\equiv\bm{q}(n\delta\tau)$ as the configuration of the system at imaginary time $n\delta\tau$. Note that motion in imaginary time is equivalent to motion in real time on the upside-down potential energy surface, $-V(\bm{q})$. \cite{Miller1971density} The resulting classical trajectory, which travels from one minimum to the other minimum is known as the instanton trajectory. Because it is the path that minimizes the action, it can be viewed as the optimal tunneling path, i.e.~it is the tunneling path that contributes the most to the path integral.

% Before discussing the higher-order corrections and the rigorous mathematical justification for instanton theory we begin by giving a simple overview of the method.  To derive instanton theory one expands the action up to second order about its minimum (within the integration domain), $\tilde{\mathbf{q}}$ (where $S'(\tilde{\bf{q}})=0$),  and then performs the resulting Gaussian integrals analytically.

Within the standard instanton approximation one stops here. However, we can do better. Equation \eqref{leading_order_inst} is just the first term in an asymptotic series in $\hbar$. Whilst it is typical to truncate the series at this point, even more accurate results can be obtained by including higher-order terms in the expansion
\begin{equation}
   \int \mathrm{d}\mathbf{q} \, e^{-S(\mathbf{q})/\hbar} \sim e^{-S(\tilde{\mathbf{q}})/\hbar}\det\left(\frac{S^{(2)}(\tilde{\mathbf{q}})}{2\pi\hbar}\right)^{-1/2}(1+a_1\hbar+\dots).
\end{equation}
The coefficients for the higher-order terms are obtained by following the procedure outlined in Appendix~\ref{Multidimensional_Asymptotics}. Simply put this involves Taylor expanding the anharmonic parts of the action and then computing the resulting integrals, which are just polynomials multiplied by a Gaussian, analytically. The integrals need only be done once for a general integral of this kind in order to obtain a general expression for the coefficients in terms of the derivatives of $S$ at $\tilde{\mathbf{q}}$. For completeness, this has been carried out in Appendix~\ref{Multidimensional_Asymptotics}. In the following we will just make use of the final equation [Eq.~\eqref{Final_Multidimensional_Integral_asymptotics}]  in order to arrive at a perturbative correction to the standard instanton expression for the tunneling splitting.

\subsection{Single-well contribution}
Defining the term that appears in the denominator of Eq.~\eqref{Delta_J=0_q_version} as
\begin{equation}
    \mathcal{G}^{(+)}(\hbar)=\int_a\mathrm{d}\bm{q}'\bra{\bm{q}'}   e^{-\tau\hat{H}/\hbar} \ket{\bm{q}_a} = A_N\int_a \mathrm{d}\mathbf{q} \, e^{-S(\mathbf{q})/\hbar}, 
\end{equation}
we can immediately write down its leading-order asymptotic approximation as 
% \begin{equation}
%     \mathcal{G}^{(+)}(\hbar) \sim \mathcal{G}^{(+)}_0(\hbar) = A_N e^{-S(\tilde{\mathbf{q}}_+)/\hbar} \det(\frac{S^{(2)}(\tilde{\mathbf{q}}_+)}{2\pi\hbar})^{-1/2}
% \end{equation}
\begin{equation}
    \mathcal{G}^{(+)}(\hbar) \sim \mathcal{G}^{(+)}_0(\hbar) = A_N e^{-S(\tilde{\mathbf{q}}_+)/\hbar} \det(\frac{\mathbf{D}^{(+)}}{2\pi\hbar})^{-1/2},
\end{equation}
where $S(\tilde{\mathbf{q}}_+)=S^{(+)}$ is the minimum of the action,  and $\mathbf{D}^{(+)}$ is the diagonal matrix of eigenvalues of the second-derivative matrix $S^{(2)}(\tilde{\mathbf{q}}_+)$. The trajectory corresponding to this term is trivial---it remains stationary at the minimum of the potential, $\bm{q}_a$. That this is the global minimum can be seen by noting that minima of the discrete Euclidean action are a compromise between minimizing the potential and minimizing the distance between successive points along the path, both of which are independently satisfied by the stationary trajectory. Note that this also means the action simplifies to $S(\tilde{\mathbf{q}}_+)=\tau V(\bm{q}_a)$, and this can be chosen to be zero by setting the zero of energy as $V(\bm{q}_a)=0$.

% Need to think about different symbol for the two stationary points $\tilde{\bf{q}}$
% \begin{equation}
% \begin{aligned}
%     \frac{\mathcal{G}^{(+)}(\hbar) }{\mathcal{G}_0^{(+)}(\hbar)}\sim 1   +\hbar&\Bigg(\frac{-3 Q_{iijj}}{4!D_{ii}D_{jj}}+\frac{9T_{iij}T_{jkk}+6T_{ijk}^2}{2!(3!)^2D_{ii}D_{jj}D_{kk}}\Bigg)+\dots
%     \end{aligned}
% \end{equation}
The higher-order terms in the asymptotic expansion can also be obtained immediately by consulting Eq.~\eqref{Final_Multidimensional_Integral_asymptotics} to give
\begin{subequations}
    \begin{equation}
    {\mathcal{G}^{(+)}(\hbar) }\sim {\mathcal{G}_0^{(+)}(\hbar)}\left( 1   +\hbar \Gamma_1^{(+)}+\mathcal{O}\left(\hbar^{2}\right)\right) ,
\end{equation}
\begin{equation}
\Gamma_1^{(+)}=\sum_{ij}\frac{-3 Q^{(+)}_{iijj}}{4!D^{(+)}_{ii}D^{(+)}_{jj}}+\sum_{ijk}\frac{9T^{(+)}_{iij}T^{(+)}_{jkk}+6[T^{(+)}_{ijk}]^2}{2!(3!)^2D^{(+)}_{ii}D^{(+)}_{jj}D^{(+)}_{kk}} ,
\end{equation}
\label{Denominator_RPI+PC}
\end{subequations}
% \begin{equation}
% \begin{aligned}
%     \frac{\mathcal{G}^{(+)}(\hbar) }{\mathcal{G}_0^{(+)}(\hbar)}\sim 1   +\hbar&\Bigg(\sum_{ij}\frac{-3 Q^{(+)}_{iijj}}{4!D^{(+)}_{ii}D^{(+)}_{jj}}\\&+\sum_{ijk}\frac{9T^{(+)}_{iij}T^{(+)}_{jkk}+6[T^{(+)}_{ijk}]^2}{2!(3!)^2D^{(+)}_{ii}D^{(+)}_{jj}D^{(+)}_{kk}}\Bigg)+\mathcal{O}\left(\hbar^{2}\right) \label{Denominator_RPI+PC}
%     \end{aligned}
% \end{equation}
where to simplify notation we have defined
\begin{subequations}
    \begin{equation}
    T^{(+)}_{ijk} = S^{(3)}_{ijk}(\tilde{\bf{q}}_+) = \sum_{\mu\nu\lambda}\frac{\partial^3 S}{\partial \tilde{q}_\mu\partial \tilde{q}_\nu\partial \tilde{q}_\lambda} C_{\mu i} C_{\nu j} C_{\lambda k} \label{T_definition} ,
    \end{equation}
    \begin{equation}
    Q^{(+)}_{iijj} = S^{(4)}_{iijj}(\tilde{\bf{q}}_+) = \sum_{\mu\nu\lambda\kappa}\frac{\partial^4 S}{\partial \tilde{q}_\mu\partial \tilde{q}_\nu\partial \tilde{q}_\lambda\tilde{q}_\kappa} C_{\mu i} C_{\nu i} C_{\lambda j} C_{\kappa j}. \label{Q_definition}
\end{equation}
\end{subequations}
Note the index of the Cartesian bead coordinates is now a combined index for the ``flattened'' array according to the definition $q_{\zeta,n}:=q_{\mu}$ with $\mu=(\zeta-1) N +n$. Additionally, to differentiate the Cartesian coordinates from the normal-mode coordinates we use Roman indices $i,j,k,l$ for the later. 

It is interesting to note that %, similarly to SCTST and VPT2,
this leading-order perturbative correction involves both third and fourth derivatives of the action, but no higher derivatives. Practically as each bead is only coupled quadratically to its neighbours, the calculation of these third and fourth derivative tensors just requires the knowledge of the third and fourth derivatives of the potential at each bead individually, similarly to SCTST and VPT2.

\subsection{Double-well (instanton) contribution}
The term that appears in the numerator of Eq.~\eqref{Delta_J=0_q_version},
\begin{equation}
    \mathcal{G}^{(-)}(\hbar)=\int_b\mathrm{d}\bm{q}'\bra{\bm{q}'}   e^{-\tau\hat{H}/\hbar} \ket{\bm{q}_a} = A_N\int_b \mathrm{d}\mathbf{q} \, e^{-S(\mathbf{q})/\hbar} ,
\end{equation}
is significantly more complicated, and requires an additional step before we can obtain its asymptotic expansion.  Firstly, note that due to the restriction on the integration domain of $\bm{q}_N$ the stationary trajectory must travel from one well to the other. 
In the limit that $\tau\to\infty$, the trajectory spends an infinite amount of time in each well, but crosses the barrier in a finite amount of time, known as the ``kink''. The additional complication arises because in this limit the action remains unchanged by changes to the time at which the kink occurs (the kink time $\eta$). This results in the minimum of  $S(\mathbf{q})$ becoming a set of continuously connected degenerate minima. However, the standard Laplace integration procedure presented above is only valid in the case that $S(\mathbf{q})$ posses a single non-degenerate minimum. It would give infinity when naively used on a problem with a continuously connected set of degenerate minima due to the presence of a zero eigenvalue of $S^{(2)}(\mathbf{q})$. We must therefore generalize the result to properly account for the continuous symmetry. Although the zero mode is also encountered in the standard (lowest order) instanton theory, here a more careful treatment is required in order to obtain the correct perturbatively corrected instanton expression.

We begin by parameterizing the set of minima using the kink time, $\eta$, 
\begin{equation}
    \tilde{\mathbf{q}}=\tilde{\mathbf{q}}(\eta).
\end{equation}
For a given $\eta$ we can then expand the action as usual to give
\begin{equation}
    S(\mathbf{q}) = S(\tilde{\mathbf{q}}(\eta)) + \frac{1}{2} (\mathbf{q}-\tilde{\mathbf{q}}(\eta))^T S^{(2)}(\tilde{\mathbf{q}}(\eta)) (\mathbf{q}-\tilde{\mathbf{q}}(\eta))  + \dots .
\end{equation}
The $\eta$-dependent matrix of second derivatives can then be diagonalized using its correspondingly $\eta$-dependent matrix of eigenvectors, $\mathbf{C}(\eta)$, as 
\begin{equation}
    \mathbf{C}^{T}(\eta) S^{(2)}(\tilde{\mathbf{q}}(\eta)) \mathbf{C}(\eta) = \mathbf{D}^{(-)},
\end{equation}
where we define the index of the zero eigenvalue such that $D^{(-)}_{ii}=0$ if and only if $i=0$. Note that, as is well known from the standard theory, in the $\tau\to\infty$ limit, the eigenvalues of $\mathbf{D}^{(-)}$ are independent of $\eta$. At each $\eta$ one can therefore define a variable transformation to a set of coordinates, $\mathbf{x}(\eta)$ that locally diagonalize $S^{(2)}(\tilde{\mathbf{q}}(\eta))$ according to the prescription
\begin{equation}
    \mathbf{x}(\eta)=\mathbf{C}^{T}(\eta) (\mathbf{q}-\tilde{\mathbf{q}}(\eta)) ,
\end{equation}
where $x_0(\eta)$ describes displacement along the tangent to the zero mode at $\eta$. We would however like to transform to a set of coordinates which include $\eta$ as a variable, to this end we define a new set of coordinates
\begin{equation}
    \bm{\xi} =  (\xi_0,\xi_1,\xi_2,...,\xi_{Nf-1}) = (\eta,x_1,x_2,...,x_{Nf-1}) \label{Variable_Transform_1}
\end{equation}
in which $x_0$ has been replaced by $\eta$, such that the original coordinates are recovered by the transformation
\begin{equation}
    q_\mu = \sum_{k=1}^{Nf-1} C_{\mu k}(\eta) x_k + \tilde{q}_\mu(\eta). \label{Variable_Transform_2}
\end{equation}
Note that when all $x_k=0$ i.e.~$k=1,Nf-1$ then $q_\mu=\tilde{q}_\mu(\eta)$.
% These coordinates correspond to defining the configuration of the path $\mathbf{q}$ in terms of displacements from some $\tilde{\mathbf{q}}(\eta)$ in directions that do not include the zero mode 

The transformation to this new set of coordinates therefore introduces a corresponding Jacobian
\begin{equation}
    \mathcal{G}^{(-)}(\hbar) =A_N \iint |J(\mathbf{x},\eta)| e^{-S(\mathbf{x},\eta)/\hbar} \mathrm{d}\mathbf{x} \mathrm{d}\eta
\end{equation}
where the integral over $\mathbf{x}$ is over only the set $x_1$,\dots$x_{Nf-1}$ which appear in $\bm{\xi}$ and (as shown in Appendix \ref{Appendix_Jacobian}) the Jacobian is given by
\begin{equation}
    J = \left|\frac{\mathrm{d} \tilde{\mathbf{q}}}{\mathrm{d} \eta}\right|-\sum_{\mu=0}^{Nf-1} \sum_{j=1}^{Nf-1} \left(\frac{\mathrm{d}^2 \tilde{q}_\mu}{\mathrm{d} \eta^2}\Bigg/\left|\frac{\mathrm{d} \tilde{\mathbf{q}}}{\mathrm{d} \eta}\right| \right)  C_{\mu j}(\eta)  x_j.\label{Jacobian_Definition}
\end{equation}
An asymptotic expansion for the inner integral can then be obtained straightforwardly using the results given in Appendix~\ref{Multidimensional_Asymptotics}. Hence, for the leading-order term we have from Eq.~\eqref{Leading_Order_Asymptotic_Multi_Dim} that
\begin{equation}
\begin{aligned}
    \mathcal{G}^{(-)}(\hbar)\sim \mathcal{G}^{(-)}_0(\hbar)&= A_N\int_0^\tau \left|\frac{\mathrm{d} \tilde{\mathbf{q}}}{\mathrm{d} \eta}\right|e^{-S(\tilde{\mathbf{q}}(\eta))/\hbar} {\rm det}'\left(\frac{\mathbf{D}^{(-)}}{2\pi\hbar}\right)^{-1/2}  \mathrm{d}\eta\\
    &= A_N \tau \left|\frac{\mathrm{d} \tilde{\mathbf{q}}}{\mathrm{d} \eta}\right|e^{-S^{(-)}/\hbar} {\rm det}'\left(\frac{\mathbf{D}^{(-)}}{2\pi\hbar}\right)^{-1/2} 
    \end{aligned}
\end{equation}
in which the prime on the determinant indicates that it is only taken over the $Nf-1$ non-zero modes, and in the second line we have integrated over the $\eta$ coordinate using the fact that in the limit that $\tau\to\infty$ then $S(\tilde{\mathbf{q}}(\eta))$,  $\left|\frac{\mathrm{d} \tilde{\mathbf{q}}}{\mathrm{d} \eta}\right|$ and $\mathbf{D}^{(-)}$ are independent of $\eta$. Note that the term
\begin{equation}
    \int_0^\tau \left|\frac{\mathrm{d} \tilde{\mathbf{q}}}{\mathrm{d} \eta}\right| \mathrm{d}\eta  = \tau \left|\frac{\mathrm{d} \tilde{\mathbf{q}}}{\mathrm{d} \eta}\right|
\end{equation}
corresponds physically to the distance in the configurational space of the path that is swept out upon changing $\eta$ i.e.~the path length.

Using the fact that the higher-order derivatives are also independent of $\eta$ in the $\tau\to\infty$ limit, we
can immediately make use of Eq.~\eqref{Final_Multidimensional_Integral_asymptotics} to write the perturbatively corrected instanton expression
\begin{subequations}
    \begin{equation}
    \mathcal{G}^{(-)}(\hbar)\sim {\mathcal{G}^{(-)}_0(\hbar)}\left(1+\hbar\Gamma_1^{(-)}+\mathcal{O}\left(\hbar^{2}\right)\right) ,
\end{equation}
\begin{equation}
\begin{aligned}
    \Gamma_1^{(-)}=&-{\sum_{ij}}'\frac{3L_iT^{(-)}_{ijj}}{3!D^{(-)}_{ii}D^{(-)}_{jj}}-{\sum_{ij}}'\frac{3Q^{(-)}_{iijj}}{4!D^{(-)}_{ii}D^{(-)}_{jj}}\\&+{\sum_{ijk}}'\frac{9T^{(-)}_{iij}T^{(-)}_{jkk}+6[T^{(-)}_{ijk}]^2}{2!(3!)^2D^{(-)}_{ii}D^{(-)}_{jj}D^{(-)}_{kk}} ,
    \end{aligned}%
\end{equation}%
% \begin{equation}
% \begin{aligned}
%     \Gamma_1^{(-)}=&-{\sum_{ij}}'\frac{3L_iT_{ijj}}{3!D_{ii}D_{jj}}-{\sum_{ij}}'\frac{3Q_{iijj}}{4!D_{ii}D_{jj}}\\&+{\sum_{ijk}}'\frac{9T_{iij}T_{jkk}+6T_{ijk}T_{ijk}}{2!(3!)^2D_{ii}D_{jj}D_{kk}}
%     \end{aligned}
% \end{equation}
\label{Numerator_RPI+PC}%
\end{subequations}%
% \begin{equation}
% \begin{aligned}
%     \frac{\mathcal{G}^{(-)}(\hbar)}{\mathcal{G}^{(-)}_0(\hbar)}&\sim 1+\hbar\Bigg(-{\sum_{ij}}'\frac{3L_iT_{ijj}}{3!D_{ii}D_{jj}}-{\sum_{ij}}'\frac{3Q_{iijj}}{4!D_{ii}D_{jj}}\\&+{\sum_{ijk}}'\frac{9T_{iij}T_{jkk}+6T_{ijk}T_{ijk}}{2!(3!)^2D_{ii}D_{jj}D_{kk}}\Bigg)+\mathcal{O}\left(\hbar^{2}\right)\Bigg)\label{Numerator_RPI+PC}
%     \end{aligned}
% \end{equation}
where the prime on the sums indicate they are only over the $j=1$ to $Nf-1$ non-zero modes. Note that as before the tensors $\mathbf{T}^{(-)}$, and $\mathbf{Q}^{(-)}$ are the third and fourth derivatives of the action, defined analogously to Eq.~\eqref{T_definition} and \eqref{Q_definition}, the only difference being that the tensors are now evaluated for the instanton path rather than the collapsed path. Additionally, we have now defined a new (rank 1) tensor $\mathbf{L}$,  
\begin{equation}
  L_i=\frac{J^{(1)}_i(\tilde{\mathbf{q}})}{J(\tilde{\mathbf{q}})} =  -\sum_{\mu=0}^{Nf-1}  \left(\frac{\mathrm{d}^2 \tilde{q}_\mu}{\mathrm{d} \eta^2}\Bigg/\left|\frac{\mathrm{d} \tilde{\mathbf{q}}}{\mathrm{d} \eta}\right|^2 \right)  C_{\mu i}(\eta),
\end{equation}
which describes how the Jacobian changes along each of the normal mode directions, the definition of which is simply obtained from  Eq.~\eqref{Jacobian_Definition}. In combination with the tensor of third derivatives this term can thus be viewed as describing perturbative corrections to the path length. Note that as the Jacobian is linear in $\mathbf{x}$ there is no quadratic fluctuation term. 
%which corresponds to the 
%The tensors $\mathbf{T}$, and $\mathbf{Q}$ are defined analogously to Eq.~\eqref{T_definition} and \eqref{Q_definition}, the only difference being that the tensors are now evaluated for the instanton path. In contrast to before we now have an additional term involving $\mathbf{L}$ which comes from the fluctuation of the Jacobian. Comparing to the definition of the Jacobian given in Eq.~\eqref{Jacobian_Definition} we have immediately that the first derivative of the Jacobian is given by

The derivatives of $\tilde{\mathbf{{q}}}$ with respect to $\eta$ describe how the beads along the path would change as the kink is acted upon by the time-translation operator.  They can be evaluated easily using finite difference as e.g.
\begin{equation}
  \frac{\partial \tilde{q}_\mu}{\partial \eta} = \frac{\partial \tilde{q}_{\zeta,n}}{\partial \eta} = \lim_{N\to\infty} \frac{\tilde{q}_{\zeta,n+1}-\tilde{q}_{\zeta,n-1}}{2\delta \tau}
\end{equation}
valid as $N\to\infty$, note for $n=N$ and $n=0$ this should be replaced with a one-sided finite difference. Similarly the second derivative is given by 
\begin{equation}
\frac{\partial^2 \tilde{q}_\mu}{\partial \eta^2} = \frac{\partial^2 \tilde{q}_{\zeta,n}}{\partial \eta^2} = \lim_{N\to\infty} \frac{\tilde{q}_{\zeta,n+1}-2\tilde{q}_{\zeta,n}+\tilde{q}_{\zeta,n-1}}{(\delta \tau)^2},
\end{equation}
which can just be set to zero for $n=0$ and $n=N$.

\subsection{Perturbatively corrected instanton theory for tunneling splittings} \label{Final_derivation}
Having obtained asymptotic expressions for both $\mathcal{G}^{(+)}(\hbar)$ and $\mathcal{G}^{(-)}(\hbar)$ we are now in a position to derive our final perturbatively corrected instanton theory expression for the tunneling splitting. Following the derivation of standard instanton theory (see e.g.~Ref.\citenum{Kleinert}) the full expression  
\begin{equation}
    \Delta(\hbar) = \lim_{\tau\to\infty} \frac{2\hbar}{\tau} \arctanh\Bigg(\frac{\mathcal{G}^{(-)}(\hbar)}{\mathcal{G}^{(+)}(\hbar)}\Bigg)
\end{equation}
can be simplified asymptotically to give
\begin{equation}
    \Delta(\hbar) \sim \lim_{\tau\to\infty}\frac{2\hbar}{\tau} \frac{\mathcal{G}_0^{(-)}(\hbar)\left(1+\hbar\Gamma^{(-)}_1+\mathcal{O}\big(\hbar^2\big)\right)}{\mathcal{G}_0^{(+)}(\hbar)\left(1+\hbar\Gamma^{(+)}_1+\mathcal{O}\left(\hbar^2\right)\right)}.
\end{equation}
The final step to obtain the asymptotic series for $\Delta$ is then simply to expand the denominator and group terms, giving
% \begin{equation}
%     \Delta(\hbar) \sim \lim_{\tau\to\infty}\frac{2\hbar}{\tau} \frac{\mathcal{G}_0^{(-)}(\hbar)}{\mathcal{G}_0^{(+)}(\hbar)}\left(1+\hbar\left(\Gamma^{(-)}_1-\Gamma^{(+)}_1\right)+\mathcal{O}\big(\hbar^2\big)\right)
% \end{equation}
\begin{subequations}
\begin{equation}
    \Delta(\hbar) \sim \Delta_\mathrm{RPI}(\hbar)\left(1+\hbar\left(\Gamma^{(-)}_1-\Gamma^{(+)}_1\right)+\mathcal{O}\big(\hbar^2\big)\right)
\end{equation}
\begin{equation}
    \Delta_\mathrm{RPI}(\hbar)= \frac{2\hbar}{\tau}\frac{\mathcal{G}_0^{(-)}(\hbar)}{\mathcal{G}_0^{(+)}(\hbar)}=\left|\frac{\mathrm{d} \tilde{\mathbf{q}}}{\mathrm{d} \eta}\right| \sqrt{\frac{2\hbar\det(\mathbf{D}^{(+)})}{\pi\,{\rm det}'\!\left({\mathbf{D}^{(-)}}\right)}} e^{-S_{\!\text{kink}}/\hbar}. \label{Final_RPI}
\end{equation}
\label{Final_Asymptotic_Expansion}%
\end{subequations}
Here $\Delta_\mathrm{RPI}(\hbar)$ is just the standard instanton expression\cite{Uses_of_Instantons,Benderskii,tunnel} for the tunneling splitting, with $S_{\!\text{kink}}=S^{(-)}-S^{(+)}$. We thus define the RPI+PC approximation to the tunneling splitting as
\begin{equation}
  \Delta_{\text{RPI+PC}}(\hbar) = \Delta_\mathrm{RPI}(\hbar)\left[1+\hbar\left(\Gamma^{(-)}_1-\Gamma^{(+)}_1\right)\right]. \label{Final_RPI+PC}
\end{equation}
This, combined with Eq.~\eqref{Final_RPI} and the definitions of $\Gamma_1^{(+)}$ and $\Gamma_1^{(-)}$ given in Eqs.~\eqref{Denominator_RPI+PC} and \eqref{Numerator_RPI+PC}, is all one needs to compute the RPI+PC approximation to the tunneling splitting.

\section{Results and Discussion}\label{Results}
\subsection{One-dimensional double well}\label{OneD_model_section}
As a first illustration of the RPI+PC method we consider the quartic double well 
\begin{equation}
    V(q) = V_0\left(\frac{q^2}{q_a^2}-1\right)^2 ,\label{1D-model}
\end{equation}
in reduced units such that $m=1$ and $\hbar=1$. For this problem, due to the particularly simple form of the potential, it is possible to obtain closed form expressions for the instanton tunneling path as well as the fluctuation factors and perturbative corrections (see e.g.~Ref.~\citenum{Kleinert}). The resulting expression for the tunneling splitting is then
\begin{equation}
    \Delta(\hbar) \sim 8\sqrt{\frac{3 \hbar V_0 S_{\!\text{kink}}}{q_a^2\pi}} \, e^{-S_{\!\text{kink}}/\hbar} \left(1-\frac{71}{72}\frac{\hbar}{S_{\!\text{kink}}}+\mathcal{O}(\hbar^2)\right)
\end{equation}
where the action is given as $S_{\!\text{kink}}=\frac{4}{3}q_a\sqrt{2V_0}$. This expression is therefore equivalent to the infinite bead limit of Eq.~\eqref{Final_Asymptotic_Expansion}, which we additionally confirmed numerically.
% and $\omega=\sqrt{8V_0}/q_0$. 

\begin{table}[t] 
\caption{Comparison of the instanton tunneling splittings $\Delta$ (in
reduced units, $m=1$ and $\hbar=1$) with the exact quantum splittings for different barrier
heights $V_0$ in the model of Eq.~\eqref{1D-model} (with a fixed harmonic frequency in the wells $\omega = \sqrt{8V_0}/q_a = \sqrt{8/}5$). Note that this means the
harmonic zero-point energy in the potential wells is approximately 0.283 (reduced units).}  
\label{Quartic_Osc_table}
\begin{ruledtabular}
\begin{tabular}{lcccc} 
$V_0$ & 2  & 1 & 0.5  & 0.25  \\ \hline
  RPI &  4.39$\times10^{-8}$ &	3.86$\times10^{-4}$	& 3.04$\times10^{-2}$ &	2.27$\times10^{-1}$          \\
  RPI+PC & 4.16$\times10^{-8}$ &	3.46$\times10^{-4}$ &	2.41$\times10^{-2}$ &	1.32$\times10^{-1}$  \\
  Exact &  4.15$\times10^{-8}$  &  3.42$\times10^{-4}$ & 2.25$\times10^{-2}$ &  1.19$\times10^{-1}$               \\
  \hline
  RPI Error & 6\% &	13\% &	35\% &	91\%  \\
  RPI+PC Error & $0.3$\% &	1\% &	7\% &	11\%   \\
\end{tabular}
\end{ruledtabular}
\end{table}
% \begin{table}[h!] 
% \caption{Comparison of the instanton tunneling splittings $\Delta$ (in
% reduced units, $\hbar=1$) with the exact quantum splittings for different barrier
% heights $V_0$ in the model of Eq.~\eqref{1D-model} (with $q_0 = 5\sqrt{
% V_0}$). Note that the
% harmonic zero-point energy in the potential wells is 0.283 (reduced units).}  
% \label{Quartic_Osc_table}
% \begin{ruledtabular}
% \begin{tabular}{lccccc} 
% $V_0$ & 2  & 1 & 0.5  & 0.25 & 0.125 \\ \hline
%   Instanton &  4.39$\times10^{-8}$ &	3.86$\times10^{-4}$	& 3.04$\times10^{-2}$ &	2.27$\times10^{-1}$   & 5.22$\times10^{-1}$       \\
%   RPI+PC & 4.16$\times10^{-8}$ &	3.46$\times10^{-4}$ &	2.41$\times10^{-2}$ &	1.32$\times10^{-1}$ & 8.53$\times10^{-2}$ \\
%   Exact &  4.15$\times10^{-8}$  &  3.42$\times10^{-4}$ & 2.25$\times10^{-2}$ &  1.19$\times10^{-1}$ & 2.46$\times10^{-1}$              \\
%   \hline
%   Inst Error & 6\% &	13\% &	35\% &	91\% & 112\% \\
%   RPI+PC Error & $0.3$\% &	1\% &	7\% &	11\% & -65\%  \\
% \end{tabular}
% \end{ruledtabular}
% \end{table}

Table \ref{Quartic_Osc_table} shows the tunneling splittings for this model for a range of different barrier heights, with $q_a$ chosen so as to keep the harmonic frequency in the wells fixed. We see that RPI+PC provides a significant improvement over the standard uncorrected RPI result at all values of the barrier height, $V_0$, considered. For the highest barrier height, $V_0=2$, equivalent to just over 7 times the harmonic zero-point energy in the well, the accuracy of RPI+PC is particularly impressive with an error of only 0.3\%. This is particularly significant as numerically exact approaches such as DMC and PIMD, struggle the most with describing small tunneling splittings, where it can be difficult to reduce the statistical error below the size of the splitting.\cite{Matyus2016tunnel1} It is well known that standard instanton theory is more accurate for high barriers than for low barriers,\cite{Kleinert,tunnel} and this remains true for RPI+PC\@. However, even for $V_0=0.25$, where the harmonic zero-point energy in the wells $\hbar\omega/2\simeq0.283$ is higher than the barrier height, the improvement over the uncorrected RPI result is significant, reducing the error from nearly a factor of 2 to just over 10\%. 
%Beyond this point the description of the gap between the ground and first excited state as a tunneling splitting becomes inappropriate 
For even lower barrier heights, the description of the gap between the ground and first excited states as a tunneling splitting ceases to be appropriate, and the problem would be better treated with other methods.\cite{Feynman,Kleinert,Cao1994CMDII,Trenins2019QCMD,Benson2020water,Fletcher2021fQCMD,Musil2022PIGS}   For example at $V_0=0.125$ (not shown in the table), where the harmonic approximation to the zero-point energy in each well is more than twice the barrier height, instanton theory is in error to the exact result by 112\% (just over a factor of two too large) and RPI+PC is no better with an error of $-$65\% (nearly a factor of three too small). 
%It is, however, important to note that the perturbative correction is still just that, perturbative, and for systems with lo..... this point  RPI+PC does not continue to give significant improvement for lower and lower barriers.  These extreme cases are not 

%[Add discussion of VPT2 - Burd/Clary results]

It is natural to wish to compare RPI+PC to other methods which involve the calculation of higher derivatives of the potential, such as VPT2 and SCTST.\cite{Miller1990SCTST} While most commonly used for the calculation of reaction rates,\cite{Wagner2013SCTST,Greene2016SCTST,Goel2018SCTST,Shan2019SCTST,Mandelli2022SCTST} recently a generalization of the SCTST and VPT2 approach has been suggested by Burd and Clary that can be used to calculate tunneling splittings.\cite{Burd2020tunnel} As they have also considered the same set of parameters for the quartic double well we can directly compare the accuracy of the methods (at least in one-dimension). Note that there is an error in the instanton results reported in Ref.~\citenum{Burd2020tunnel}, however it does not affect the qualitative comparison. Their VPT2 approach is the most similar to RPI+PC in that it requires the calculation of third and fourth derivatives of the potential, albeit only at the saddle point. While VPT2 shows an improvement over the uncorrected RPI tunneling splittings for the lowest barrier height, $V_0=0.25$, it is less accurate than uncorrected RPI theory at higher values of $V_0$. Here we find that, with RPI+PC, including third and fourth derivatives leads to results that are significantly more accurate than VPT2 at all values of $V_0$ considered, and in fact also more accurate than VPT4, which requires fifth and sixth derivatives of the potential. %, albeit only at the minimum and saddle point.

% This is reflected in the results with the most significant errors seen 

%This is reflected in the results with the most significant errors seen at the lowest $V_0=0.125$ where the harmonic approximation to the zero-point energy in each well is more than twice the barrier height, and the smallest error at large. [this is not really a tunneling splitting anymore...and the problem would be better treated with other methods...]

% ground state wavefunction is higher than the barrier height  the largest errors se  As expected we see that both the original instanton theory and the RPI+PC result are more accurate for higher barriers. The RPI+PC results show a significant over improvement for 

%The error made by instanton theory can be understood qualitatively as arising from the assumption that in the full path integral fluctuations 

% table of results for different V0s - this should ideally go to point where method starts to break

% plot vs $\hbar$ for fixed V0 - probably not necessary as I assume it's similar to studying changing V0 with fixed hbar. - maybe do this for 2D model instead

% give Kleinert's 
% analytic formula for quartic double well

\subsection{Two-dimensional model}\label{TwoD_model_section}

%Here as the problem is only one dimensional the error made by instanton theory is due to approximating fluctuations along the path rather than perpendicular to it. Although the instanton path itself captures the full anharmonicity of the potential the approximation to the path integral assumes that the fluctuation of each point along the path  
In one dimension the error made by instanton theory is due solely to the harmonic approximation made to the fluctuations along the path.  However, in multiple dimensions, instanton theory makes the additional approximation that fluctuations perpendicular to the instanton path are harmonic. In contrast, instanton theory with perturbative corrections is able to capture the effect of anharmonicity in the perpendicular coordinates. In order to illustrate this in a system for which it is still straightforward to obtain exact quantum results for comparison we consider a simple two dimensional model. 

\begin{figure}[t]
    \centering
    \includegraphics{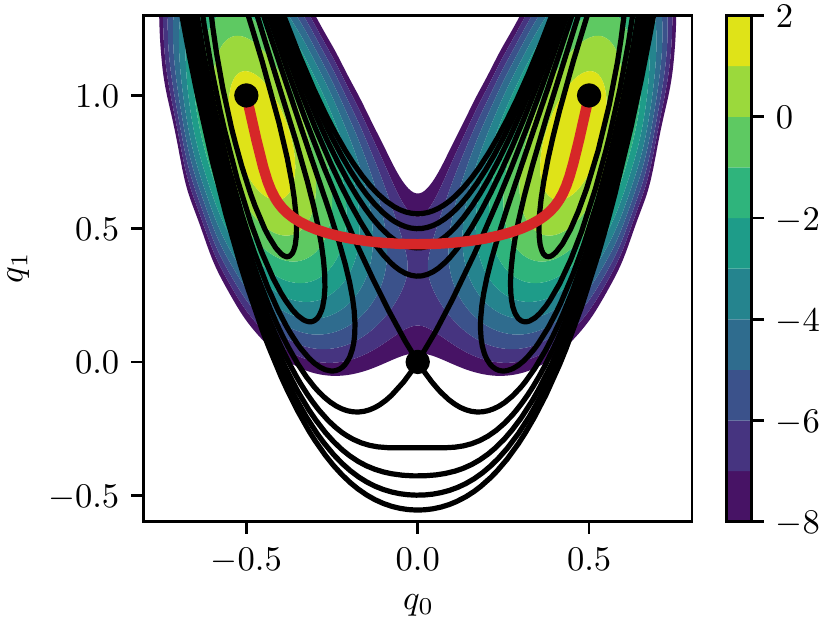}
    \caption{Illustration of the two-dimensional model from Sec.~\ref{TwoD_model_section} with $\alpha=400$. Black contours show the potential energy surface with a spacing of 25 reduced energy units, the instanton path is shown as a red line, the stationary points are shown as black circles, and the colour map shows the natural logarithm of the ground-state density.  }
    \label{fig:2D}
\end{figure}

The model we consider for this purpose is a quartic double well coupled to a quartic oscillator, %adapted from Ref.~\citenum{Benderskii},
\begin{equation}
    V(\bm{q}) = -aq_0^2 + bq_0^4 + \frac{1}{2} \omega^2 \left[ q_1 - \frac{cq_0^2}{\omega^2} \right]^2 + \alpha \left[ q_1 - \frac{cq_0^2}{\omega^2} \right]^4, \label{eq:2DPES}
\end{equation}
% \begin{equation}
%     V(q_0,q_1) = -a_0q_0^2 + b_0q_0^4 + a_1 \left[ q_1 - \frac{cq_0^2}{2\alpha} \right]^2 + b_1 \left[ q_1 - \frac{cq_0^2}{m\omega^2} \right]^4, 
% \end{equation}
where the parameters, in reduced units ($\hbar=1$ and $m=1$), are given by $a=800$, $b=1600$, $c=1600$, $\omega=20$, and $\alpha$ (the parameter that controls the anharmonicity in the $q_1$ coordinate) is varied between 0 and $400$. The minima and the saddle point of the potential are located at $(\pm1/2,1)$ and $(0,0)$ respectively independent of the choice of $\alpha$. The potential difference between the minima and saddle point is $100$ reduced units, and is also independent of $\alpha$. Figure \ref{fig:2D} shows a contour plot of the model (with $\alpha=400$) along with the corresponding instanton path and a colour map showing the natural logarithm of the ground-state density. Both the instanton path and the ground-state wavefunction demonstrate characteristic corner-cutting behaviour, preferring to take a shorter path through the barrier via regions of high potential energy, rather than following the minimum-energy path through the saddle point.

%Figure \ref{fig:2Danharmonicity} shows the behaviour of the tunneling splitting as a function of the anharmonicity, $\alpha$, in the $q_1$ coordinate. We see that the exact result increases by 25\% as the anharmonicity is increased from $\alpha=0$ to $\alpha=800$.

\begin{figure}[t]
    \centering
    \includegraphics{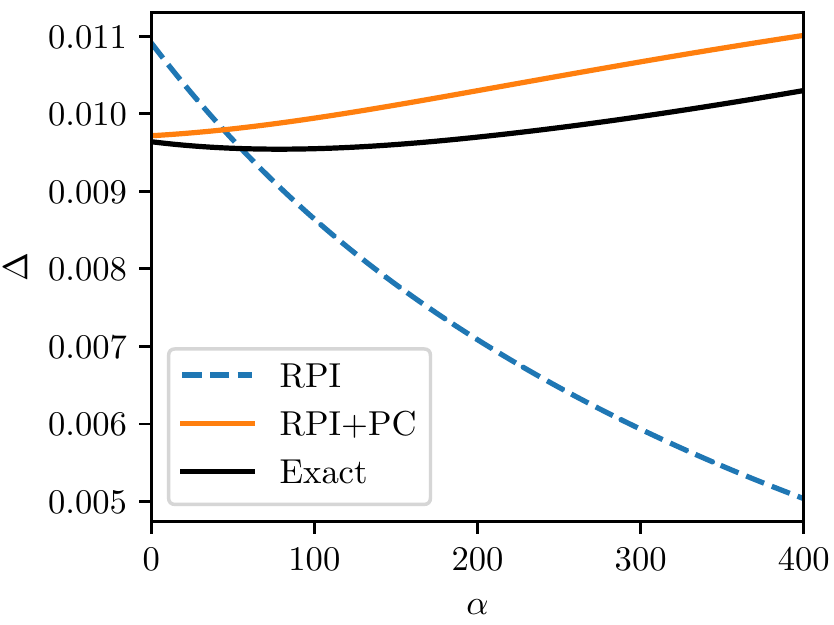}
    \caption{Plot showing the tunneling splitting $\Delta$ (in reduced units) as a function of the anharmonicity in the $q_1$ coordinate, $\alpha$, (also in reduced units) for the two dimensional model [Eq.~\eqref{eq:2DPES}]. The exact results are calculated using a two-dimensional DVR.}
    \label{fig:2Danharmonicity}
\end{figure}

Figure \ref{fig:2Danharmonicity} shows the behaviour of the tunneling splitting as a function of the anharmonicity, $\alpha$, in the $q_1$ coordinate. While the RPI+PC and exact results are in good agreement for all values of $\alpha$ considered, the dependence on $\alpha$ predicted by standard instanton theory (RPI) is qualitatively wrong. Before discussing the RPI+PC results it is instructive to first analyse the error of the standard instanton result.
% This behaviour can be understood as follows.
At $\alpha=0$, the error in the uncorrected RPI result is dominated by anharmonicity along the path, i.e.~the low-barrier problem, resulting in a value of $\Delta$ which is slightly too large with respect to the exact result. This can be understood by noting that when $\alpha=0$ the potential is entirely harmonic in the $q_1$ direction and hence perpendicular fluctuations near the center of the path are described exactly by the standard instanton result. The results are thus analogous to the one-dimensional quartic double well of Sec.~\ref{OneD_model_section} at $V_0\approx1$: the tunneling splitting is approximately $10^{-4}$ times the barrier height and the errors in the uncorrected RPI result and RPI+PC are 13\% and 1\% respectively. For larger values of $\alpha$ the anharmonicity perpendicular to the path begins to become important, leading to a decrease in the standard instanton result. For a small range of $\alpha$, near $\alpha\approx50$, the uncorrected RPI result is fortuitously accurate due to a cancellation of errors. However, as the anharmonicity continues to increase the  uncorrected RPI result becomes less and less accurate, such that, at $\alpha=400$, the standard instanton result is just over a factor of 2 too small. Here making a harmonic approximation around the instanton path makes the potential along the $q_1$ coordinate in the barrier region appear narrower than reality. This leads to an underestimation of the extent of the fluctuations around the path and results in a prediction of $\Delta$ that is too small.
%Here making a harmonic approximation around the instanton path makes the barrier region appear higher and narrower than reality, resulting in an underprediction of $\Delta$. 

%Except in the small region where the errors made by the standard instanton approach approximately cancel we see that the RPI+PC result is more accurate than the standard instanton theory. 

%...Punchy sentence to begin that highlights how RPI+PC is much better than instanton and captures anharmonicity perpendicular to the path....

%Figure~\ref{fig:2Danharmonicity} clearly illustrates that RPI+PC is not only able to correct errors in the standard instanton approach that arise from low barriers, but also that it is able to correctly capture the effect of anharmonicity perpendicular to the path.
RPI+PC goes beyond the standard instanton result by including the effect of anharmonicity perpendicular to the path, and hence is able to correctly capture the increase in the tunneling splitting with increasing $\alpha$. 
%We see that RPI+PC successfully corrects the error of the standard instanton result, and captures the effects of both anharmonicity along the path and anharmonicity perpendicular to the path.  
Even at the largest value of the anharmonicity shown in Fig.~\ref{fig:2Danharmonicity}, where the uncorrected RPI result is in error by more than a factor of two, RPI+PC has an error of only 7\%.  Given the success of standard instanton theory it seems unlikely that larger values of $\alpha$ are representative of tunneling splittings in most chemical systems.\cite{tunnel,water,tropolone} However, it is not uncommon in the calculation of reaction rates for there to be significant anharmonicity.\cite{Muonium,Gao2021muonium,Truhlar,JoeFaraday,Faraday2022RateGeneralDisc}
%It is thus encouraging to note that even well beyond this range at $\alpha=800$ where the standard instanton is more than a factor of four too small the RPI+PC result continues to be within a few percent of the exact result.
 For more significantly anharmonic systems one might expect that a simple linear perturbative correction will cease to be sufficient, and that it would be necessary to make use of approximate resummation techniques. The applicability of these techniques to chemical problems is an interesting area of future research.

\subsection{Full-dimensional malonaldehyde}
\begin{figure}[t]
    \centering
    \includegraphics{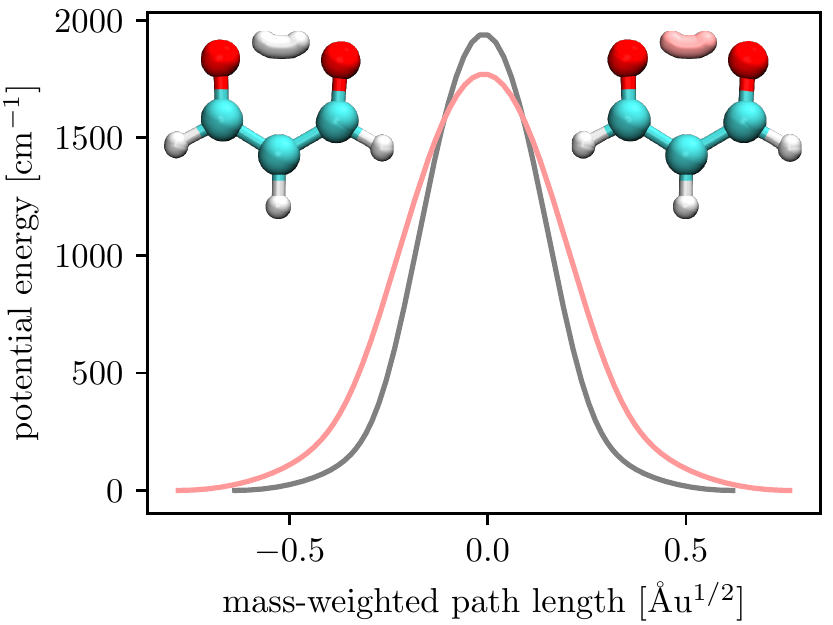}
    \caption{Potentials along the instanton pathways (insets left and right) for malonaldehyde (grey) and deuterated malonaldehyde (pink).}
    \label{fig:malonaldehyde}
\end{figure}

Finally we illustrate the applicability of the RPI+PC approach to real  systems by considering the prototypical molecular example of tunneling splitting, malonaldehyde. This system has become a ``go-to'' system for illustrating the accuracy of new methods, both for fitting potential energy surfaces,\cite{Wang2008malonaldehydePES,Mizukami2014malonaldehyde} and for computing tunneling splittings.\cite{Milnikov2001,Milnikov2003,Schroeder2011malonaldehyde,Wang2013Qim,Matyus2016tunnel1} The potential energy surface we consider here is the 2014 PES developed by Mizukami, Habershon and Tew.\cite{Mizukami2014malonaldehyde} The potential was fit to electronic-structure calculations at the CCSD(T)(F12*) level, with a root-mean-square accuracy of $\sigma=0.3$\% in the relative energy range, and additionally outputs analytical gradients and second derivatives.\footnote{We note that as the fitting function is analytic it would in principle be possible to also obtain analytical third and fourth derivatives, however we found that using finite difference presented no difficulties.} This is the most accurate potential energy surface for malonaldehyde developed to date, and has the advantage that there are exact results for comparison. This includes pre-existing DMC\cite{Mizukami2014malonaldehyde} and PIMD\cite{Matyus2016tunnel1,Vaillant2018dimer} results, as well as new Smolyak-grid-based wavefunction results from the ElVibRot code.\cite{Lauvergnat2023Malonaldehyde} We note that while MCTDH calculations have been performed on malonaldehyde\cite{Hammer2011malonaldehyde,Schroeder2011malonaldehyde} so far this has only been for the 2008 PES of Wang \emph{et al.}\cite{Wang2008malonaldehydePES} %From the point as well as having analytical 2nd derivatives. 
In the following we consider both the regular (fully hydrogenated) malonaldehyde as well as one of its deuterated isotopologues, in which the transferring hydrogen is replaced by deuterium.

Figure~\ref{fig:malonaldehyde} shows the potential energy along the instanton path for both hydrogen and deuterium transfers, with the corresponding geometric representations as insets. In both cases there is significant corner cutting, with the highest potential significantly above the energy of the saddle point (1411~cm$^{-1}$).\cite{Mizukami2014malonaldehyde} As expected, the corner cutting is less pronounced in the case of deuterium transfer as the heavier mass of deuterium makes the barrier in mass-weighted coordinates wider.

\begin{table}[t] 
\caption{Comparison of the tunneling splittings $\Delta$ for malonaldehyde and its deuterated isotopologue.  We find the uncorrected RPI result is converged to within 0.5\% at 2048 beads with $T=\hbar/k_B\tau=50$~K for both H and D. The RPI+PC correction factor, $\Delta_{\text{RPI+PC}}(\hbar)/\Delta_\mathrm{RPI}(\hbar)$, is converged to less than 0.005 (i.e.~0.5\%) at $N=256$ beads for H and D. The statistical errors for the PIMD, and experimental results are quoted as 1 standard error. The error of the DMC result is a combination of the statistical error and an estimate of the lower bound of the error due to the fixed node approximation (0.3~cm$^{-1}$), for more details see Ref.~\citenum{Mizukami2014malonaldehyde}. The ElVibRot results are computed using a Smolyak grid based wavefunction method, and the estimated errors reflect incomplete basis set convergence.\cite{Lauvergnat2023Malonaldehyde}}  
\label{Full_Malonaldehyde}
\begin{ruledtabular}
\begin{tabular}{l|cccc} 
 & H (cm$^{-1}$)  & D (cm$^{-1}$)  \\ \hline
  PIMD Ref.~\citenum{Matyus2016tunnel1} &  20.6 $\pm$ 1.3   &  $-$            \\
  PIMD Ref.~\citenum{Vaillant2018dimer} &  19.3 $\pm$ 0.2   &  $-$            \\
  DMC\cite{Mizukami2014malonaldehyde} &  21.0 $\pm$ 0.4   & 3.2 $\pm$ 0.4             \\
  ElVibRot\cite{Lauvergnat2023Malonaldehyde} & 21.7 $\pm$ 0.3 & 2.9 $\pm$ 0.1 \\
  RPI &  19.25     & 2.67 \\
  RPI+PC & 22.1  & 2.96 
 \\ \hline
  Experiment\cite{Baughcum1984malonaldehyde,Turner1984malonaldehyde,Firth1991malonaldehyde,Baba1999malonaldehyde} & 21.583  & 2.915 $\pm$ 0.004

\end{tabular}
\end{ruledtabular}
\end{table}

Table \ref{Full_Malonaldehyde} shows the tunneling splittings for both malonaldehyde and its deuterated isotopologue. Before discussing the accuracy of the instanton and RPI+PC results, it is necessary to first comment on the results from the fully quantum methods: PIMD, DMC and the ElVibRot calculations. The first thing one notes is that, in the case of H transfer, each of these ``formally exact'' methods predict significantly different values for the tunneling splitting. For DMC it is relatively clear what the errors are. In addition to its statistical error, there is an additional systematic error due to the requirement to approximate the nodal plane of the first excited state wavefunction. The authors of Ref.~\citenum{Mizukami2014malonaldehyde} state that the error caused by this fixed node approximation is difficult to estimate, but they give a lower bound of 0.3~cm$^{-1}$ and include this in their estimated error bars. While the large statistical error bars of the PIMD results from Ref.~\citenum{Matyus2016tunnel1} are consistent with all of the other results, the discrepancy between the PIMD results of Ref.~\citenum{Vaillant2018dimer} and the other methods is harder to explain. The authors of Ref.~\citenum{Vaillant2018dimer} attempted to address this, considering the possibility of contamination from rotationally excited tunneling splittings, as well as acknowledging that they cannot rule out sampling errors from non-ergodicity. In either case, given the large difference between these results and the DMC and ElVibRot calculation, we think that it is likely that there exists some systematic error in this PIMD calculation. In contrast, the ElVibRot calculations, which are performed using a wavefunction method based on a Smolyak grid, are not expected to suffer from significant systematic error.  Note that the error bars given for the ElVibRot calculation are an estimate of the error due to incomplete basis set size.\cite{Lauvergnat2023Malonaldehyde} It is also notable that the ElVibRot results obtain much closer agreement with the experimental measurements. We will therefore use this as our benchmark.
%[ElVibRot close to expt]

%Firstly we note that although both the DMC and PIMD results are formally exact, they each have rather large errors. In the case of the PIMD calculation these errors are predominantly statistical. (It should be noted that this was one of the first calculations using PIMD and more recent developments have reduced the level of statistical noise) However

The RPI+PC results show a clear improvement over the uncorrected RPI results. In the case of H transfer the error relative to the ElVibRot  reduces from about $-$11\% to 2\% and the the error relative to the experimental results from $-$11\% to 2.4\%. In the case of D transfer the error reduces from around $-$8\% to be within the error bars of the ElVibRot calculation and within 2\% of the experimental result. Given that there is such a large spread in the supposedly exact results, a sceptical reader might be concerned---despite the discussion of the previous paragraph---that we have simply compared to the exact results that best favour RPI+PC\@. To this end we give a more detailed discussion of the error one would expect from our method. Firstly we note that the RPI+PC tunneling splitting is larger than the uncorrected RPI result for both the H and D transfers. We see a 15\% increase in the tunneling splitting in going from RPI to RPI+PC for the H transfer and an 11\% increase in the tunneling splitting for the D transfer. By analogy with the two dimension model considered in Sec.~\ref{TwoD_model_section} this indicates that the dominant error in the uncorrected RPI result likely arises from ignoring anharmonicity perpendicular to the path, rather than from the barrier being low.  Assuming that the behaviour seen in the two dimensional model is transferable to the full-dimensional malonaldehyde then we would expect that the RPI+PC result would have slightly over corrected the tunneling splitting. This is exactly what we observe when we make the assumption that the ElVibRot calculations are the most reliable benchmark. We note that, the close agreement of RPI+PC and ElVibRot with the experimental measurements indicates that the main error in the uncorrected RPI result was the instanton approximation and not the PES.

%Paragraph pointing out that supposedly exact methods are not up to much, and one might as well use RPI+PC which is both much cheaper, easier to use and more reliable.
The exponential scaling of the nuclear Hilbert space means that, even with modern compression techniques for reducing the storage required, such as used in the ElVibRot code, the size of system to which grid-based methods can be applied is limited. In principle PIMD and DMC appear to offer an attractive alternative as they are fully quantum mechanical and avoid this exponential scaling. However, as the results of Table~\ref{Full_Malonaldehyde} show there is apparently a significant risk of systematic errors in both approaches. Even without this, as DMC and PIMD are statistical methods, they require a large number of potential evaluations in order to reach convergence. The instanton and RPI+PC calculations on the other hand only require knowledge of the potential in a small region around the instanton path. Perhaps more importantly both instanton theory and RPI+PC are simple to use. Finding the instanton path is equivalent to finding the minimum of the path-integral action, and once this is done the RPI+PC expression is simple to apply. There are no adjustable parameters and no need for technical input to choose basis sets, nodal planes or paths for thermodynamic integration. RPI+PC thus stands out as a computationally efficient, accurate and reliable method for calculating tunneling splittings.

Tables showing the convergence of the uncorrected RPI splitting as well as the RPI+PC correction factor, $\Delta_{\text{RPI+PC}}/\Delta_\mathrm{RPI}$, are given in the supplementary information. We find that for both H and D transfer the uncorrected RPI result is converged to within 0.5\% with $N=2048$ beads at an imaginary time corresponding to $T=\hbar/k_B\tau=50$~K, and that the RPI+PC correction factor is converged to less than 0.005 (i.e., corresponding to a $0.5$\% correction of the instanton rate) at only $N=256$ beads. Given the cost of evaluating third and fourth derivatives, it is particularly encouraging that the RPI+PC correction factor converges significantly faster than the leading-order term. This indicates that when combined with techniques for reducing the computational cost, such as using beads spaced equally in position (rather than time)\cite{Milnikov2001,Rommel2011grids,Cvitas2018instanton} and by interpolating the derivatives along the path, that RPI+PC can be made a very cost efficient method.

% Calculation details:
% Number of beads, temperature required

% convergence of the standard instanton vs correction factor

% Details given in the supplementary information

% ease with which the

% the mht2014 PES has analytic derivatives (at least to 2nd order) \cite{Mizukami2014malonaldehyde}
% and there are DMC results for comparison
% also \cite{Matyus2016tunnel1} does malonaldehyde with PIMD.
% At least mention MCTDH \cite{Hammer2011malonaldehyde,Schroeder2011malonaldehyde} although they used different PES from \cite{Wang2008malonaldehydePES}.

% [notes: I claimed that our malonaldehyde calculates were 10\% out \cite{tropolone}]
% [best estimate of barrier height of malonaldehyde is 1429 cm-1 from fc-CCSD(T)(F12*)/def2-QZVPP//fc-CCSD(T)(F12*)/def2-TZVPP,\cite{Mizukami2014malonaldehyde} what is it for our PES? Is it 1411cm-1?  This might explain why we are slightly overpredicting the splitting - but check using alpha scaling

% $\alpha$ scaling:\\
% we have $\alpha=1.0015$, so that would make the corrected potential 1413cm-1 - [hmm, not really want I hoped for, but could still be ok as we're corner cutting anyway, so information on TS is sort of unreliable]
% Corrected results:

% D: 2.907\\
% H: 21.768
% ]

\section{Conclusions and future work} \label{Conclusion}
In this paper we have demonstrated how RPI+PC can be applied to calculate accurate tunneling splittings in molecular systems. By considering a series of model systems we have shown that RPI+PC improves upon the predictions of standard instanton theory, both in systems with low barriers and in systems with significant anharmonicity perpendicular to the tunneling path. Additionally, using an accurate full-dimensional potential energy surface, we have applied RPI+PC to calculate the ground-state tunneling splitting in malonaldehyde. Comparison with experiment and new benchmark results on this potential shows that, not only does RPI+PC have an error of 2\% or less for both hydrogen and deuterium transfer, but also that this error is smaller than those of previous calculations performed with formally ``exact'' methods, DMC and PIMD\@. This highlights a great advantage of the RPI+PC approach, its simplicity. While PIMD and DMC are fully quantum mechanical, both are sensitive to exactly how the calculations are performed. In contrast, having found the instanton path, the RPI+PC result can be calculated analytically using only information about the potential along the instanton path.

One of the key advantages of instanton theory is its computational efficiency, in particular, that it does not require knowledge of the global potential energy surface. This allows instanton theory to be combined directly with ab-initio calculations of the potential,\cite{Milnikov2003,chiral} or with schemes to learn the region of the potential local to the instanton.\cite{GPR,TransferLearning} In this work we have demonstrated RPI+PC can be efficiently applied with precomputed potential energy surfaces. For future applications with on-the-fly ab-initio potentials, it will be interesting to explore approaches that further reduce the cost of the RPI+PC method, especially because the calculation of third and fourth derivatives is expensive. However we note that this is routine in VPT2 for calculating anharmonic corrections to spectra.\cite{Franke2022HowToVPT2} Furthermore, recent advances in the use of machine learning may help to significantly reduce the computational cost. \cite{GPR,TransferLearning,Bac2022HigherDerivatives} Using these and similar interpolation techniques may be possible to reduce the number of third and fourth derivatives needed. This could potentially be achieved by interpolating the elements of these tensors along the path, in analogy with existing approaches for reducing the number of hessian evaluations.\cite{Muonium} This may mean the tensors only need to be evaluated at a few points along the path, making the cost comparable to a SCTST calculation. It may also be possible to take only certain elements of the tensors (treating anharmonicity only in selected modes), or to provide simple low-barrier corrections by only taking the terms along the tunneling path, in analogy to suggestions from Burd and Clary.\cite{Burd2020tunnel} 

With these further developments we believe RPI+PC can become the ``go to'' method for computing accurate tunneling splittings. There are a number of systems to which it will be immediately interesting to apply the theory. For example the water dimer and trimer, where the barrier between the minima is relatively low and for which the uncorrected RPI result was found to be in error by as much as a factor of two.\cite{water} Due to the presence of multiple wells these systems have more complicated tunneling splitting patterns than malonaldehyde. In spite of this additional complication the present approach generalizes straightforwardly to these situations. We note however that for the water dimer, as well as other systems which have high rotational temperatures and significant rovibrational coupling, it may be necessary to further refine the current treatment of the rotational modes. In a similar vein it will also be interesting to explore the calculation of rotationally excited tunneling splittings. So far this has been only been tackled with instanton theory by numerical integration over Euler angles via the calculation of a large number of paths,\cite{Vaillant2018instanton} although it should be possible to develop an asymptotic approach which would significantly simplify the calculation of such splittings.

%Promise of ab-initio don't need global PES

%Advantage is that in principle it does not need global potential energy surface. In the present work we have applied the method to a precomputed potential energy surface, however in the future to make use of 

%As we have demonstrated here the RPI+PC is already straightforward to apply to molecular systems for which a PES has been fitted. However for direct application with ab-initio electronic structure it will likely be necessary to  

In addition to the calculation of tunneling splittings, there are exciting prospects for the generalization of the RPI+PC approach to add anharmonic corrections to instanton rate theory. Instanton 
 rate theory is a powerful tool for including the effect of tunneling on reaction rates, that can be used in both electronically adiabatic\cite{Miller1975semiclassical,AdiabaticGreens} and electronically non-adiabatic systems.\cite{GoldenGreens,PhilTransA,inverted,thiophosgene,nitrene} %Just like in the case of tunneling splittings it is based on picking out the dominant tunneling path in the path-integral description of the reaction rate. It is therefore able to capture important effects such as corner cutting that simple one dimensional theories struggle to accurately describe. In the Born-Oppenheimer limit instanton rate theory can be viewed as an extension of Eyring transition state theory. Just like Eyring-transition state theory instanton, and the instanton approach to tunneling splittings described in this work it therefore treats  vibrations harmonically. %however the standard theory cannot describe anharmonicity perpendicular to the tunneling path.
 While the mathematical techniques needed to derive such a correction will be the same as those in this paper, the different structure of the rate problem will require further theoretical development. For example, as the effect of anharmonicity can be more pronounced in reaction rates than for tunneling splittings, we expect it will be necessary to develop schemes to approximately resum the asymptotic series. Such techniques are standard in theoretical physics,\cite{PeskinSchroeder} and have already been used in the context of electronically non-adiabatic reaction rates when considering the perturbative expansion in diabatic coupling.\cite{4thorder} Because there are many systems in which the effects of anharmonicity and tunneling can both be significant, the ability to include  anharmonic corrections in instanton rate calculations would be very powerful. % method. %would lead to a very powerful method.
 At present one must typically choose between accurately describing the tunneling path using instanton theory, including the effects of anharmonicity with approaches such as SCTST\cite{Miller1990SCTST,Mandelli2022SCTST} or using more expensive path-integral calculations on the full potential energy surface.\cite{Miller2003QI,RPMDrate,RPMDrefinedRate,Suleimanov2013RPMDrate,Lawrence2020rates} A generalization of the RPI+PC approach to reaction rates promises to have the advantages of all of these approaches. 

\section*{Supplementary}
Supplementary material is available showing the convergence of the results.
%(probably not needed)
%if anything, then just convergence tests.  Could also present results from double fixed ends vs single fixed end.

\section*{Data Availability Statement}
The data that supports the findings of this study are available within the article and its supplementary material.

\section*{Conflict of Interest}
The authors have no conflicts to disclose.

\section*{Author Contributions}
{\bf Joseph E. Lawrence:} Conceptualization (equal); Methodology (lead);  Software (supporting); Validation (equal); Formal Analysis (equal); Investigation (equal); Data Curation (supporting); Writing - Original Draft Preparation (lead); Writing - Review \& Editing (equal); Visualization (equal); Supervision (equal).
{\bf Jind\v{r}ich Du\v{s}ek:} Methodology (supporting); Software (lead); Validation (equal);  Formal Analysis (equal); Investigation (equal); Data Curation (lead); Writing - Review \& Editing (support); Visualization (equal). 
{\bf Jeremy O. Richardson:} Conceptualization (equal); Writing - Review \& Editing (equal); Visualization (lead); Supervision (equal).
%Conceptualization, JEL and JOR; Methodology, JEL; Software, JD and JEL; Validation, JEL and JD; Formal Analysis, JEL and JD; Investigation, JEL and JD; Data Curation, JD and JEL; Writing - Original Draft Preparation, JEL; Writing - Review \& Editing, JEL, JOR and JD; Visualization, JOR, JD and JEL; Supervision, JEL and JOR; 

\section*{Acknowledgements}
We would like to thank David Lauvergnat for sharing the ElVibRot results for malonaldehyde with us ahead of publication. JEL was supported by an ETH Z\"urich Postdoctoral Fellowship and JD by a scholarship from the Bakala Foundation. 

\appendix
%\section*{Appendix}
\section{Avoiding explicitly integrating over Euler angles} % $\bm{\Omega}$}
\label{Appendix_Euler_angles}
Before we consider a simple procedure for avoiding the explicit integration over $\bm{\Omega}$, we first observe that the original expression for the tunneling splitting in systems with no rotational symmetry can be reformulated by integrating the final position over a finite region around  the minimum $b$ i.e.~by considering
\begin{equation}
\begin{aligned}
    \int_b \mathrm{d} \bm{q}' \mel{\bm{q}'}{ e^{-\tau\hat{H}/\hbar} }{\bm{q}_a} &=\int_b \mathrm{d} \bm{q}'  \braket{\bm{q}'}{\psi_{+}} \braket{\psi_{+}}{\bm{q}_a} e^{-\tau E_{+}/\hbar}\\
&+\int_b \mathrm{d} \bm{q}' \braket{\bm{q}'}{\psi_{-}} \braket{\psi_{-}}{\bm{q}_a} e^{-\tau E_{-}/\hbar}.
    \end{aligned}
\end{equation}
This can again be simplified using the antisymmetry of the wavefunctions to give an expression involving integrals over the symmetrically related region around the minimum $a$ as
\begin{equation}
\begin{aligned}
    \int_b \mathrm{d} \bm{q}' \mel{\bm{q}'}{ e^{-\tau\hat{H}/\hbar} }{\bm{q}_a} &=\int_a \mathrm{d} \bm{q}'  \braket{\bm{q}'}{\psi_{+}} \braket{\psi_{+}}{\bm{q}_a} e^{-\tau E_{+}/\hbar}\\
&-\int_a \mathrm{d} \bm{q}' \braket{\bm{q}'}{\psi_{-}} \braket{\psi_{-}}{\bm{q}_a} e^{-\tau E_{-}/\hbar}.
    \end{aligned}
\end{equation}
The derivation then follows as before with a new definition of the quantities $\rho$ and $\gamma$
\begin{equation}
    \int_a \mathrm{d} \bm{q}'  \braket{\bm{q}'}{\psi_{\pm}} \braket{\psi_{\pm}}{\bm{q}_a}=\rho e^{\pm\gamma}.
\end{equation}
Hence we arrive at a slightly different (but also formally exact) expression for the tunneling splitting for systems without rotational degrees of freedom as
\begin{equation}
    \Delta = \lim_{\tau\to\infty}\frac{2\hbar}{\tau}\arctanh\left(\frac{\int_b \mathrm{d}\bm{q}'\mel{\bm{q}'}{ e^{-\tau\hat{H}/\hbar} }{\bm{q}_a}}{\int_a \mathrm{d}\bm{q}'\mel{\bm{q}'}{  e^{-\tau\hat{H}/\hbar} }{\bm{q}_a}}\right).
\end{equation}
Equivalently, by introducing projection operators onto the $a$ and $b$ regions, for example $\hat{P}_a=\theta(-s(\hat{\bm{q}}))$ and $\hat{P}_b=\theta(s(\hat{\bm{q}}))$, where $\theta(x)$ is the Heaviside step function and $s(\bm{q})=0$ is the dividing surface which separates the regions corresponding to the two minima we can also write this as\footnote{Note that in principle we do not need the regions to obey $\hat{P}_a+\hat{P}_b=1$, and we choose this here just for simplicity of notation.}
\begin{equation}
    \Delta = \lim_{\tau\to\infty}\frac{2\hbar}{\tau}\arctanh\left(\frac{\int \mathrm{d}\bm{q}'\mel{\bm{q}'}{ \hat{P}_b e^{-\tau\hat{H}/\hbar} }{\bm{q}_a}}{\int\mathrm{d}\bm{q}'\mel{\bm{q}'}{ \hat{P}_a  e^{-\tau\hat{H}/\hbar} }{\bm{q}_a}}\right).
\end{equation}

This idea can be generalized for systems with rotational symmetry to simplify the expression for the $J=0$ tunneling splitting. The arguments above generalize straightforwardly to allow us to write
\begin{equation}
    \Delta = \lim_{\tau\to\infty}\frac{2\hbar}{\tau}\arctanh\left(\frac{\int \mathrm{d}\bm{q}'\mel{\bm{q}'}{ \hat{P}_b\hat{P}_{J=0} e^{-\tau\hat{H}/\hbar} }{\bm{q}_a}}{\int\mathrm{d}\bm{q}'\mel{\bm{q}'}{ \hat{P}_a \hat{P}_{J=0}  e^{-\tau\hat{H}/\hbar} }{\bm{q}_a}}\right).
\end{equation}
It is then intuitive to see that the projection onto $J=0$ can be removed, given that $\hat{P}_{J=0}$ and $\hat{P}_b$ commute, and that $\int \mathrm{d}\bm{q}'\bra{\bm{q}'}\hat{P}_{J=0}=\int \mathrm{d}\bm{q}'\bra{\bm{q}'}$. Alternatively, one can show this directly with a couple of simple manipulations, the first of which is to act $\hat{P}_b$ on $\hat{P}_{J=0}$ and then integrate over the outer position integral
\begin{equation}
\begin{aligned}
    \int\mathrm{d}\bm{q}'\bra{\bm{q}'} \hat{P}_{b}\hat{P}_{J=0} &=  \frac{1}{8\pi^2}\iiint\mathrm{d}\bm{q}'  \mathrm{d}\bm{q}\mathrm{d}\bm{\Omega}\mel{\bm{q}'}{\hat{P}_{b}   }{R_{\bm{\Omega}} (\bm{q})}\bra{\bm{q}} \\
    &=   \frac{1}{8\pi^2} \int \mathrm{d}\bm{q} \int \mathrm{d}  {\bm{\Omega}} \, \theta(s(R_{\bm{\Omega}} (\bm{q})))\bra{\bm{q}}.
    \end{aligned}
\end{equation}
Then, noting that $\theta(s(R_{\bm{\Omega}} (\bm{q})))$ is independent of the choice of ${\bm{\Omega}}$ we can replace $\theta(s(R_{\bm{\Omega}} (\bm{q})))$ with $\theta(s(\bm{q}))$ and integrate over ${\bm{\Omega}}$ to give the desired result
\begin{equation}
\begin{aligned}
    \int\mathrm{d}\bm{q}'\bra{\bm{q}'} \hat{P}_{b}\hat{P}_{J=0} =    \int \mathrm{d}\bm{q}  \, \theta(s(\bm{q}))\bra{\bm{q}}=\int\mathrm{d}\bm{q}'\bra{\bm{q}'} \hat{P}_{b}.
    \end{aligned}
\end{equation}
The same argument clearly applies replacing $\hat{P}_b$ with $\hat{P}_a$ and hence we arrive at a simplified expression for the tunneling splitting
\begin{equation}
    \Delta = \lim_{\tau\to\infty}\frac{2\hbar}{\tau}\arctanh\left(\frac{\int \mathrm{d}\bm{q}'\mel{\bm{q}'}{\hat{P}_b e^{-\tau\hat{H}/\hbar}}{\bm{q}_a}}{\int\mathrm{d}\bm{q}'\mel{\bm{q}'}{\hat{P}_a   e^{-\tau\hat{H}/\hbar}}{\bm{q}_a}}\right).
\end{equation}
We thus keep one end fixed but allow the other to be flexible.  This avoids the complication of having both ends free (which has an overall rotational degree of freedom with more zero modes to worry about). Note that inserting the projection onto rotational states in this location only works for $J=0$, the derivation for the equivalent expression for other rotational projections is slightly more involved.

% \subsection{Projections onto other rotational states}
% The projection onto other rotational states can be written as
% \begin{equation}
%     \hat{P}_{J,K,M} = \frac{1}{8\pi^2} \int \mathrm{d}\bm{q} \int \mathrm{d} \Omega \, D_{J,K,M}(\Omega) \ketbra{R_\Omega (\bm{q})}{\bm{q}}
% \end{equation}
% note that we cannot insert this projection next to the integral over all space for any states other than $J=0$ as all other states do not transform as the totally symmetric irrep and hence just integrate to zero. Instead we insert the projection on the right of the propagator giving
% \begin{equation}
%     \Delta_{J,K,M} = \lim_{\tau\to\infty}\frac{2\hbar}{\tau}\arctanh\left(\frac{\int \mathrm{d}\bm{q}'\bra{\bm{q}'} \hat{P}_b e^{-\tau\hat{H}/\hbar}\hat{P}_{J,K,M} \ket{\bm{q}_a}}{\int\mathrm{d}\bm{q}'\bra{\bm{q}'} \hat{P}_a   e^{-\tau\hat{H}/\hbar} \hat{P}_{J,K,M}\ket{\bm{q}_a}}\right).
% \end{equation}

% \begin{equation}
%     \Delta_{J,K,M} = \lim_{\tau\to\infty}\frac{2\hbar}{\tau}\arctanh\left(\frac{\int\mathrm{d}\Omega\int  \mathrm{d}\bm{q}'\bra{\bm{q}'} \hat{P}_b e^{-\tau\hat{H}/\hbar}D_{J,K,M}(\Omega) \ket{R_{\Omega}(\bm{q}_a)}}{\int\mathrm{d}\Omega\int \mathrm{d}\bm{q}'\bra{\bm{q}'} \hat{P}_a   e^{-\tau\hat{H}/\hbar} D_{J,K,M}(\Omega)\ket{R_{\Omega}(\bm{q}_a)}}\right).
% \end{equation}

%Basic ideas of Z- approach

\section{Introduction to asymptotic analysis} \label{Appendix:Intro_to_asymptotics}
For readers less familiar with the techniques of asymptotic analysis we give a brief introduction to the key concepts used in the paper. 
Firstly we introduce the idea asymptotic equivalence, denoted $\sim$, which generalizes the usual equality, $=$. If we have two functions $A(\epsilon)$ and $B(\epsilon)$ we can say 
\begin{equation}
    A(\epsilon)\sim B(\epsilon) \text{ as } \epsilon\to0 
\end{equation}
read as ``$A$ is asymptotic to $B$ as $\epsilon$ tends to zero'' which is equivalent to the statement
\begin{equation}
    \lim_{\epsilon\to0}
\frac{A(\epsilon)}{ B(\epsilon) } =1.
\end{equation}
Asymptotic equivalence $\sim$ shares many properties with the more usual equality, $=$, however there are some important differences which we now highlight. Firstly, as is clear from the definition above, asymptotic equivalence is only meaningful when specified along with an asymptotic parameter which approaches some limit (here $\epsilon$, but usually $\hbar$ in the main part of the paper). Secondly, although multiplying both sides of an asymptotic relation by a constant leaves the relation intact, in general this is not true if one adds or subtracts arbitrary functions from each side of the relation. For a complete description of the properties of asymptotic equivalence see for example Ref.~\citenum{BenderBook}.

Whilst asymptotic equivalence is a useful concept on its own, it becomes especially useful in allowing us to define the idea of an asymptotic series. Asymptotic series generalize the idea of convergent series expansions (such as Taylor series) to obtain representations of functions in terms of series which are not necessarily convergent. They are particularly useful as a way of developing accurate approximations for functions which cannot be written as a simple analytic expression.  As an example, if a function $I(\epsilon)$ has an asymptotic series representation
\begin{equation}
  I(\epsilon) \sim e^{-b/\epsilon}\epsilon^m\sum_{n=0}^{\infty} a_n\epsilon^n \text{ as } \epsilon\to0,
\end{equation}
(where $m$ may be some non-integer power) then for each term in the series we have (for any $N$)
\begin{equation}
    \left(I(\epsilon) - e^{-b/\epsilon}\epsilon^m\sum_{n=0}^{N} a_n\epsilon^n \right) \sim e^{-b/\epsilon}\epsilon^m a_{N+1}\epsilon^{N+1}
\end{equation}
(as $\epsilon\to0$). As this example illustrates asymptotic series need not necessarily be simple power series expansions and may also contain other functions of the asymptotic parameter, such as exponentials and logarithms. The standard instanton theory expression for $\Delta(\hbar)$ is the leading-order term in an asymptotic series in $\hbar$, i.e.~the term involving $a_0$. The aim of the present paper is to develop a method for calculating the term corresponding to $a_1$.   

When one is used to dealing with convergent series representations such as Taylor series, the fact that asymptotic series need not necessarily be convergent may appear rather worrisome. However, divergent asymptotic series are actually often much more powerful than their convergent counterparts. In particular whilst convergent Taylor series often require the summation of a large number of terms before accurate results are obtained, asymptotic series are often very accurate after the inclusion of only one or two terms. Since the series is ultimately divergent one may worry that it is not clear when one should stop including terms. Fortunately, as a general rule one can expect to obtain optimal results by truncating the asymptotic series just before the smallest term.\cite{BenderBook} Hence, while the terms continue to decrease in magnitude, we can expect to get more accurate results by including additional terms in the series. %Furthemore, there are also other methods of obtaining information about the result from the asymptotic series. Such methods include Borel resummation, Pad\' e  approximation or analytic continuation and we refer to them jointly as resummation techniques. While such techniques can greatly improve the accuracy of the result, choosing the best resummation technique is a difficult problem.

\subsection{One-dimensional integrals}\label{Appendix:One_dim_asymptotics}
Whilst we have now explained the concept of an asymptotic series, we have not yet discussed how one can actually obtain such a series. For the present work we need only consider how to obtain asymptotic expansion of integrals such as appear in the path-integral expression for the tunneling splitting. Hence, as an illustrative example consider a one-dimensional integral
\begin{equation}
I(\epsilon) = \int_{a}^b g(x) e^{-f(x)/\epsilon} \mathrm{d}x,
\end{equation}
where $f(x)$ plays the role of the action, to which we would like to obtain an asymptotic expansion as $\epsilon\to0$. 
To do so, we argue that the integral is dominated by the region around the minimum of $f(x)$ in $[a,b]$, which we will denote $\tilde{x}$ such that $f^{(1)}(\tilde{x})=0$.\footnote{It will be sufficient for the present work to just consider the case where the minimum appears in the interior of the integration region and not at its boundary.} Hence, provided $g(\tilde{x})\neq0$,\footnote{The generalization to the case where $g(\tilde{x})=0$ is straightforward but is not needed here.} then as  $\epsilon\to0$ the integrand looks approximately like a Gaussian
\begin{equation}
    g(x) e^{-f(x)/\epsilon} \approx g(\tilde{x}) e^{-f(\tilde{x})/\epsilon-(x-\tilde{x})^2f^{(2)}(\tilde{x})/2\epsilon}
\end{equation}
where to maintain consistency with later sections which will involve higher-order derivatives we use $f^{(n)}(\tilde{x})$ to denote the $n^{\rm th}$ derivative of $f$ at $\tilde{x}$. To see why this is a good approximation note that as $\epsilon\to0$ the functional form of the exponent does not change but rather just becomes scaled by a large constant term $1/\epsilon$. Hence the domain in $x$ over which the exponent is well approximated by a parabola is unchanged, however the absolute value of the exponent at the point where the higher-order terms become important becomes larger and larger as $\epsilon\to0$. Therefore the integral becomes dominated by a narrower and narrower region over which it becomes better and better approximated by a Gaussian. This idea is formalized by Laplace's method %Laplace's method (no need for complex plane here).
 which says that the integral is asymptotic to
\begin{equation}
\begin{aligned}
    I(\epsilon) &\sim  \int_{-\infty}^{\infty} g(\tilde{x}) e^{-f(\tilde{x})/\epsilon-(x-\tilde{x})^2f^{(2)}(\tilde{x})/2\epsilon}\mathrm{d}x\\
    &=g(\tilde{x}) e^{-f(\tilde{x})/\epsilon}\sqrt{\frac{2\pi\epsilon}{f^{(2)}(\tilde{x})}}. 
    \end{aligned}
\end{equation}
Note that the limits can be taken to infinity because as $\epsilon\to0$ and the peak gets narrower and narrower the contribution from the regions outside the integration range become negligible. A rigorous proof of this asymptotic relation can be found in e.g.~Ref.~\citenum{BenderBook}. 

This is just the first term in the asymptotic expansion. To obtain higher-order terms in the expansion we begin by making a change of variables from $x$ to $u=(x-\tilde{x})/\sqrt{\epsilon}$ which will allow us to more easily identify the $\epsilon$ dependence of different terms in the integral by allowing the $\epsilon$ dependence to be factored out of the integrals. One then simply Taylor expands both 
\begin{equation}
    g(x)= \sum_{n=0}^\infty\frac{g^{(n)}(\tilde{x})u^n\epsilon^{n/2}}{n!} \label{1D_g_taylor_series}
\end{equation}
as well as the remaining part of the exponential
% \begin{equation}
%     \frac{e^{-f(x)/\epsilon}}{e^{-f(\tilde{x})/\epsilon}} = e^{-{u^2f^{(2)}(\tilde{x})}/{2}}\sum_{l=0}^\infty\frac{1}{l!}\left(\sum_{m=3}^\infty-\frac{u^m\epsilon^{m/2}f^{(m)}(\tilde{x})}{m!\epsilon}  \right)^l
% \end{equation}
\begin{equation}
    \frac{e^{-f(x)/\epsilon}}{e^{-f(\tilde{x})/\epsilon-{u^2f^{(2)}(\tilde{x})}/{2}}}  = \sum_{l=0}^\infty\frac{1}{l!}\left(\sum_{m=3}^\infty-\frac{u^m\epsilon^{(m-2)/2}f^{(m)}(\tilde{x})}{m!}  \right)^l. \label{1D_exponential_taylor_series}
\end{equation}
Each term in the asymptotic series can then be found by grouping terms with the same order of $\epsilon$, and noting that odd powers of $u$ will integrate to zero. 

To obtain the first-order correction we thus need only retain terms up to order $\epsilon$ in both Eq.~\eqref{1D_g_taylor_series} and Eq.~\eqref{1D_exponential_taylor_series} corresponding to
% \begin{equation}
%     g(x)= g(\tilde{x}) + g^{(1)}(\tilde{x})u\epsilon^{1/2} + \frac{g^{(2)}(\tilde{x})u^2\epsilon}{2}+\dots
% \end{equation}
\begin{equation}
    \frac{g(x)}{g(\tilde{x})}= 1 + \frac{g^{(1)}(\tilde{x})u\epsilon^{1/2}}{g(\tilde{x})} + \frac{g^{(2)}(\tilde{x})u^2\epsilon}{2!g(\tilde{x})}+\mathcal{O}\left(\epsilon^{3/2}\right)
\end{equation}
and
% \begin{equation}
% \begin{aligned}
%     \frac{e^{-f(x)/\epsilon}}{e^{-f(\tilde{x})/\epsilon}} = e^{-{u^2f^{(2)}(\tilde{x})}/{2}}\Bigg(&1-\frac{f^{(3)}u^3\epsilon^{1/2}}{3!}\\&-\left(\frac{f^{(4)}u^4}{4!}-\frac{[f^{(3)}]^2u^6}{2!(3!)^2}\right)\epsilon+\dots\Bigg)
%     \end{aligned}
% \end{equation}
% \begin{equation}
% \begin{aligned}
%     \frac{e^{-f(x)/\epsilon}}{e^{-f(\tilde{x})/\epsilon}} = e^{-{u^2f^{(2)}(\tilde{x})}/{2}}\Bigg(&1-\frac{f^{(3)}(\tilde{x})u^3\epsilon^{1/2}}{3!}-\frac{f^{(4)}(\tilde{x})u^4\epsilon}{4!}\\&+\frac{[f^{(3)}(\tilde{x})]^2u^6\epsilon}{2!(3!)^2}+\dots\Bigg)
%        \end{aligned}
% \end{equation}
\begin{equation}
\begin{aligned}
    \frac{e^{-f(x)/\epsilon}}{e^{-f(\tilde{x})/\epsilon-{u^2f^{(2)}(\tilde{x})}/{2}}} = &1-\frac{f^{(3)}(\tilde{x})u^3\epsilon^{1/2}}{3!}-\frac{f^{(4)}(\tilde{x})u^4\epsilon}{4!}\\&+\frac{[f^{(3)}(\tilde{x})]^2u^6\epsilon}{2!(3!)^2}+\mathcal{O}\left(\epsilon^{3/2}\right).
       \end{aligned}
\end{equation}
% Hence the 1st term in the series (beyond the leading-order term) is given by
% \begin{equation}
%     g(x)\frac{e^{-f(x)/\epsilon}}{e^{-f(\tilde{x})/\epsilon}} = e^{-{u^2f^{(2)}(\tilde{x})}/{2}}(1+cu\epsilon^{n/2}+\epsilon\left( \frac{g^{(2)}(\tilde{x})u^2}{2}-\frac{u^4f^{(4)}(\tilde{x})}{4!}-\frac{u^4g^{(1)}(\tilde{x})f^{(3)}(\tilde{x})}{3!}+\frac{u^6[f^{(3)}(\tilde{x})]^2}{2!(3!)^2}+\dots\right)
% \end{equation}
% \begin{equation}
%     \frac{g(x) e^{-f(x)/\epsilon}}{g(\tilde{x})e^{-f(\tilde{x})/\epsilon-(x-\tilde{x})^2f^{(2)}(\tilde{x})/2\epsilon}} = \left(\sum_{n=0}^\infty\frac{g^{(n)}(\tilde{x})(x-\tilde{x})^n}{n!}\right)\dots 
% \end{equation}
% \begin{equation}
%     e^{-f(x)/\epsilon} = \exp(-\frac{f(\tilde{x})}{\epsilon}-\frac{(x-\tilde{x})^2f^{(2)}(\tilde{x})}{2\epsilon})\sum_{l=0}^\infty\left(\sum_{m=3}^\infty-\frac{(x-\tilde{x})^mf^{(m)}(\tilde{x})}{m!\epsilon}  \right)^l
% \end{equation}
Dividing out the leading-order term,
\begin{equation}
\begin{aligned}
    I_0(\epsilon) =g(\tilde{x}) e^{-f(\tilde{x})/\epsilon}\sqrt{\frac{2\pi\epsilon}{f^{(2)}(\tilde{x})}},
    \end{aligned}
\end{equation}
we can then write the asymptotic series as
% in terms of expectation values over the Gaussian distribution corresponding to the leading-order term as
\begin{equation}
\begin{aligned}
    I(\epsilon) \sim I_0(\epsilon)\Bigg\langle &1+\frac{g^{(2)}(\tilde{x})u^2\epsilon}{2!g(\tilde{x})}-\frac{g^{(1)}(\tilde{x})f^{(3)}(\tilde{x})u^4\epsilon}{3!g(\tilde{x})}\\ &-\frac{f^{(4)}(\tilde{x})u^4\epsilon}{4!}+\frac{[f^{(3)}(\tilde{x})]^2u^6\epsilon}{2!(3!)^2}+\mathcal{O}\left(\epsilon^{2}\right)\Bigg\rangle
    \end{aligned}
\end{equation}
where the expectation values are defined as
% \begin{equation}
%   \langle u^n\rangle = \frac{\int_{-\infty}^{\infty} u^n  e^{-u^2f^{(2)}(\tilde{x})/2}\mathrm{d}u}{\int_{-\infty}^{\infty}  e^{-u^2f^{(2)}(\tilde{x})/2}\mathrm{d}u} = [f^{(2)}(\tilde{x})]^{-n/2} (n-1)!!
% \end{equation}
\begin{equation}
  \langle u^n\rangle = \frac{\int_{-\infty}^{\infty} u^n  e^{-u^2f^{(2)}(\tilde{x})/2}\mathrm{d}u}{\int_{-\infty}^{\infty}  e^{-u^2f^{(2)}(\tilde{x})/2}\mathrm{d}u} .
\end{equation}
Note that all expectation values involving odd powers are zero and those with even powers  are given by
\begin{equation}
  \langle u^n\rangle =  [f^{(2)}(\tilde{x})]^{-n/2} \frac{n!}{(n/2)!2^{n/2}}.
\end{equation}
Hence one obtains the first-order term in the asymptotic series as
\begin{equation}
\begin{aligned}
    I(\epsilon) \sim I_0(\epsilon)\Bigg( &1+\frac{g^{(2)}(\tilde{x})\epsilon}{2g(\tilde{x})f^{(2)}(\tilde{x})}-\frac{g^{(1)}(\tilde{x})f^{(3)}(\tilde{x})\epsilon}{2g(\tilde{x})[f^{(2)}(\tilde{x})]^{2}}\\ &-\frac{f^{(4)}(\tilde{x})\epsilon}{8[f^{(2)}(\tilde{x})]^{2}}+\frac{5[f^{(3)}(\tilde{x})]^2\epsilon}{24[f^{(2)}(\tilde{x})]^{3}}+\mathcal{O}\left(\epsilon^{2}\right)\Bigg).
    \end{aligned}
\end{equation}
Note that the resulting first-order correction involves both the third derivative as well as the fourth derivative of the exponent. This is reminiscent of other perturbative theories such as VPT2 in which both the third and fourth derivatives of the potential appear in the leading-order correction. Additionally the fluctuation of the prefactor $g(x)$ is now incorporated via its first and second derivatives. 

\subsection{Multi-dimensional integrals}\label{Multidimensional_Asymptotics}
 In order to obtain an asymptotic expansion of a path integral it is necessary to discuss the generalization of the one-dimensional theory discussed above to multiple dimensions. Hence consider the following integral
\begin{equation}
    I(\epsilon) = \int  g(\bm{x})e^{-f(\bm{x})/\epsilon} \mathrm{d}\bm{x}.
\end{equation}
Just as in the one-dimensional case the leading-order term in the asymptotic expansion is found by taking a Gaussian approximation to the integrand giving
\begin{equation}
    I(\epsilon) \sim \int g(\tilde{\bm{x}})e^{-f(\tilde{\bm{x}})/\epsilon - \sum_{\mu\nu}({x}_\mu-\tilde{{x}}_\mu)f^{(2)}_{\mu\nu}(\tilde{\bm{x}})({x}_\nu-\tilde{{x}}_\nu)/2\epsilon} \mathrm{d}\bm{x},
\end{equation}
which upon performing the Gaussian integrals allows us to write
\begin{equation}
    I(\epsilon) \sim I_0(\epsilon) =  g(\tilde{\bm{x}}) e^{-f(\tilde{\bm{x}})/\epsilon} \det(\frac{f^{(2)}(\tilde{\bm{x}})}{2\pi\epsilon})^{-1/2}, \label{Leading_Order_Asymptotic_Multi_Dim}
\end{equation}
where the main difference is now that the second derivative $f^{(2)}(\tilde{\bm{x}})$ is a matrix.

To obtain higher-order terms in the asymptotic series it is necessary to transform to the basis which diagonalizes the second-derivative matrix
\begin{equation}
    \mathbf{C}^{T} f^{(2)}(\bm{\tilde{x}}) \mathbf{C} = \mathbf{D}
\end{equation}
where $\mathbf{C}$ is the matrix of normalized eigenvectors and $\mathbf{D}$ is a diagonal matrix of the corresponding eigenvalues.  Hence making the variable transformation
\begin{equation}
    \epsilon^{-1/2} \mathbf{C}^{T} (\bm{x}-\tilde{\bm{x}}) =  \bm{u}
\end{equation}
the exponent simplifies to
\begin{equation}
    \sum_{\mu\nu}\frac{1}{2\epsilon}({x}_\mu-\tilde{{x}}_\mu)f^{(2)}_{\mu\nu}(\tilde{\bm{x}})({x}_\nu-\tilde{{x}}_\nu) = \sum_i\frac{1}{2} u_i^2 D_{ii}.  %=  \sum_i u_i^2 \lambda_i 
\end{equation}
Note we will use Roman indices ($i$, $j$, $k$ \dots) to refer to quantities in the diagonal basis and Greek indices for quantities in the original basis.
Following the same steps as in the one-dimensional case  one then obtains %(using the Einstein summation convention)
\begin{equation}
\begin{aligned}
    \frac{I(\epsilon)}{I_0(\epsilon)} \sim \Bigg\langle &1+\epsilon\Bigg(\frac{\sum_{ij}g^{(2)}_{ij}(\tilde{\bm{x}})u_iu_j}{2g(\tilde{\bm{x}})}-\frac{\sum_{ijkl}g^{(1)}_i(\tilde{\bm{x}})f^{(3)}_{jkl}(\tilde{\bm{x}})u_iu_ju_ku_l}{3!g(\tilde{\bm{x}})}\\ &-\frac{\sum_{ijkl}f^{(4)}_{ijkl}(\tilde{\bm{x}})u_iu_ju_ku_l}{4!}+\frac{[\sum_{ijk}f^{(3)}_{ijk}(\tilde{\bm{x}})u_iu_ju_k]^2}{2!(3!)^2}\Bigg)\\&+\mathcal{O}\left(\epsilon^{2}\right)\Bigg\rangle,
    \end{aligned}
\end{equation}
where the derivatives are now tensors defined in the diagonal basis according to
\begin{equation}
    f^{(n)}_{ij\dots}(\tilde{\bm{x}}) = \sum_{\mu\nu\dots}C_{\mu i} C_{\nu j}\dots\frac{\partial}{\partial \tilde{x}_\mu}\frac{\partial}{\partial \tilde{x}_\nu}\dots f(\tilde{\bm{x}}) 
\end{equation}
with $g^{(n)}(\tilde{\bm{x}})$ defined similarly, and the expectation value is defined as
\begin{equation}
    \langle \cdots \rangle = \frac{\int \cdots  e^{-\sum_i u^2_i D_{ii}/2} \mathrm{d}\bm{u}}{\int   e^{-\sum_i u^2_i D_{ii}/2} \mathrm{d}\bm{u}}.
\end{equation}

To obtain the final expression for the higher-order corrections we need to evaluate the expectation values. This is straightforward using Wick's theorem, which states that for a multivariate Gaussian expectation value with an even number of terms $n$ 
\begin{equation}
    \langle u_{i_{1}} u_{i_{2}} \dots u_{i_{n}} \rangle = \sum_{P} \langle u_{k_{1}} u_{k_{2}} \rangle \dots \langle u_{k_{n-1}} u_{k_{n}} \rangle
\end{equation}
where the sum on the right hand side is performed over all possible pairings, $P$, of the indices. Note that by symmetry expectation values with an odd number of terms are zero, and also since we are in the diagonal basis
\begin{equation}
    \langle u_i u_j \rangle = \langle u_i u_j \rangle \delta_{ij}.
\end{equation}
Hence combining these results along with the symmetry of the derivative tensors under permutation of indices we have that
\begin{subequations}
 \begin{equation}
\begin{aligned}
  \sum_{ijkl}g^{(1)}_i(\tilde{\bm{x}})f^{(3)}_{jkl}(\tilde{\bm{x}})\langle u_iu_ju_ku_l\rangle =\sum_{ij} 3 g^{(1)}_i(\tilde{\bm{x}})f^{(3)}_{ijj}(\tilde{\bm{x}})\langle u_i^2\rangle \langle u_j^2\rangle
    \end{aligned}
\end{equation}
% \begin{equation}
% \begin{aligned}
%   g^{(1)}_i(\tilde{\bm{x}})f^{(3)}_{jkl}(\tilde{\bm{x}})\langle u_iu_ju_ku_l\rangle = 3 g^{(1)}_i(\tilde{\bm{x}})f^{(3)}_{ijj}(\tilde{\bm{x}})\langle u_i^2\rangle \langle u_j^2\rangle
%     \end{aligned}
% \end{equation}
\begin{equation}
    \sum_{ijkl}f^{(4)}_{ijkl}(\tilde{\bm{x}})\langle u_iu_ju_ku_l\rangle =\sum_{ij} 3f^{(4)}_{iijj}(\tilde{\bm{x}})\langle u_i^2\rangle \langle u_j^2\rangle
\end{equation}
\begin{equation}
\begin{aligned}
    \bigg\langle \bigg[\sum_{ijk}f^{(3)}_{ijk}(\tilde{\bm{x}})u_iu_ju_k\bigg]^2 \bigg\rangle &=\sum_{ijk} 9 f^{(3)}_{iij}(\tilde{\bm{x}})f^{(3)}_{jkk}(\tilde{\bm{x}})\langle u_i^2\rangle\langle u_j^2\rangle\langle u_k^2 \rangle\\&+\sum_{ijk}6f^{(3)}_{ijk}(\tilde{\bm{x}})f^{(3)}_{ijk}(\tilde{\bm{x}})\langle u_i^2\rangle\langle u_j^2\rangle\langle u_k^2 \rangle.
    \end{aligned}
\end{equation}
\end{subequations}
These can be simplified using
\begin{equation}
    \langle u_i^2\rangle = \frac{1}{D_{ii}}
\end{equation}
to obtain the final expression for the asymptotic series up to order $\epsilon$
\begin{equation}
\begin{aligned}
    \frac{I(\epsilon)}{I_0(\epsilon)} \sim \Bigg( &1+\epsilon\Bigg(\sum_i\frac{g^{(2)}_{ii}(\tilde{\bm{x}})}{2g(\tilde{\bm{x}})D_{ii}}-\sum_{ij}\frac{3g^{(1)}_i(\tilde{\bm{x}})f^{(3)}_{ijj}(\tilde{\bm{x}})}{3!g(\tilde{\bm{x}})D_{ii}D_{jj}}\\&-\sum_{ij}\frac{3f^{(4)}_{iijj}(\tilde{\bm{x}})}{4!D_{ii}D_{jj}}+\sum_{ijk}\frac{9f^{(3)}_{iij}(\tilde{\bm{x}})f^{(3)}_{jkk}(\tilde{\bm{x}})}{2!(3!)^2D_{ii}D_{jj}D_{kk}}\\&+\sum_{ijk}\frac{6f^{(3)}_{ijk}(\tilde{\bm{x}})f^{(3)}_{ijk}(\tilde{\bm{x}})}{2!(3!)^2D_{ii}D_{jj}D_{kk}}\Bigg)+\mathcal{O}\left(\epsilon^{2}\right)\Bigg) .\label{Final_Multidimensional_Integral_asymptotics}
    \end{aligned}
\end{equation}
This is the key expression that is required to derive the perturbatively corrected instanton expression for the tunneling splitting.

\section{Derivation of the Jacobian} \label{Appendix_Jacobian}
In the following we derive the Jacobian for the variable transformation defined by Eqs.~\eqref{Variable_Transform_1} and \eqref{Variable_Transform_2}. For the uncorrected RPI expression one just needs the value of the Jacobian at the instanton geometry, which is easy to obtain with simple qualitative arguments. In order to compute the perturbative corrections, however, one needs to know how the Jacobian fluctuates around the instanton path. Hence, a more careful analysis is required. We note that the variable transformation made here - and hence the also the mathematical steps used - are closely related to that used in other contexts, for example in the derivation of the reaction path Hamiltonian.\cite{Miller1980RPH} The derivation of the perturbative corrections for the quartic double well given by Kleinert in Ref.~\citenum{Kleinert} also involve the computation of Jacobian fluctuations, the principle difference is that here we require expressions in discrete rather than continuous form. Additionally our considerations are general to any potential (rather than just the one-dimensional quartic double well considered by Kleinert). 

Starting from the definition of the variable transformation in Eqs.~\eqref{Variable_Transform_1} and \eqref{Variable_Transform_2} we can see that
\begin{equation}
\begin{aligned}
    J &= \left| \begin{matrix} \frac{\partial q_0}{\partial \eta} & \frac{\partial q_0}{\partial x_1} & \frac{\partial q_0}{\partial x_2} & \dots \\
    \frac{\partial q_1}{\partial \eta} & \frac{\partial q_1}{\partial x_1} & \frac{\partial q_1}{\partial x_2} & \dots \\
    \vdots & \vdots & \vdots & \ddots
\end{matrix} \right| \\
& = \left| \begin{matrix} \frac{\partial q_0}{\partial \eta} & C_{01}(\eta) & C_{02}(\eta) & \dots \\
    \frac{\partial q_1}{\partial \eta} & C_{11}(\eta) & C_{12}(\eta) & \dots \\
    \vdots & \vdots & \vdots & \ddots
\end{matrix} \right|
\end{aligned}
\end{equation}
This can then be simplified using Cramer's rule, which says that the solution to a system of linear equations 
\begin{equation}
    \mathbf{C} \bm{a} = \bm{b}
\end{equation}
is given by
\begin{equation}
    a_j = \frac{\det(\mathbf{C}_j)}{\det(\mathbf{C})}
\end{equation}
where $\mathbf{C}_j$ is the matrix $\mathbf{C}$ with its $j^{\rm th}$ column replaced by the vector $\bm{b}$. Hence identifying $\bm{b}=\frac{\partial \mathbf{q}}{\partial \eta}$ we have that
\begin{equation}
    J =\det(\mathbf{C}_0) =  \left(\mathbf{C}^{-1}\frac{\partial \mathbf{q}}{\partial \eta}\right)_0 \det(\mathbf{C}).
\end{equation}
Then using the fact that $\mathbf{C}$ is an orthogonal matrix we can simplify this further to give
\begin{equation}
    J =  \left(\mathbf{C}^{T}\frac{\partial \mathbf{q}}{\partial \eta}\right)_0 =  \sum_{\mu=0}^{Nf-1} C_{\mu0} \frac{\partial q_\mu}{\partial \eta}.
\end{equation}
In its current form the $\mathbf{x}$ dependence of the Jacobian is hidden within the derivative of the original coordinates. The dependence on $\mathbf{x}$ can be made explicit by directly differentiating Eq.~\eqref{Variable_Transform_2} to obtain
\begin{equation}
    \frac{\partial q_\mu}{\partial \eta} = \sum_{k=1}^{Nf-1} \frac{\partial C_{\mu k}(\eta)}{\partial \eta} x_k + \frac{\partial \tilde{q}_\mu}{\partial \eta}
\end{equation}
and hence the Jacobian can be written as
\begin{equation}
    J =  \sum_{\mu=0}^{Nf-1} \sum_{k=1}^{Nf-1} C_{\mu0}(\eta)  \frac{\partial C_{\mu k}(\eta)}{\partial \eta} x_k + \sum_{\mu=0}^{Nf-1} C_{\mu 0}(\eta)\frac{\partial \tilde{q}_\mu}{\partial \eta}\label{Jacobian_expanded}
\end{equation}
where now all $\mathbf{x}$ dependence can be explicitly seen.

The $\mathbf{x}$ independent term in the Jacobian can be simplified using the following two results. Firstly, noting that since $\tilde{\mathbf{q}}(\eta)$ does not depend on $\mathbf{x}$ we have
\begin{equation}
    \frac{\partial \tilde{q}_\mu}{\partial \eta} = \frac{\mathrm{d} \tilde{q}_\mu}{\mathrm{d} \eta}.
\end{equation}
Secondly, note that the eigenvector corresponding to the zero mode of the Hessian in the original coordinate frame must point tangentially in the direction of the continuous symmetry i.e.
\begin{equation}
    C_{\mu0}(\eta) = \frac{\mathrm{d} \tilde{q}_\mu}{\mathrm{d} \eta}\Bigg/\left|\frac{\mathrm{d} \tilde{\mathbf{q}}}{\mathrm{d} \eta}\right|. \label{Zero_mode_eigenvector}
\end{equation}
Hence inserting these results into Eq.~\eqref{Jacobian_expanded} we have that 
\begin{equation}
    J =  \sum_{\mu=0}^{Nf-1} \sum_{k=1}^{Nf-1} C_{\mu0}(\eta)  \frac{\partial C_{\mu k}(\eta)}{\partial \eta} x_k + \left|\frac{\mathrm{d} \tilde{\mathbf{q}}}{\mathrm{d} \eta}\right|.
\end{equation}

The linear term in $\mathbf{x}$ in the Jacobian can also be simplified. To do so we begin by noting that the orthogonality of $\mathbf{C}$ implies
\begin{equation}
    \sum_{\mu=0}^{Nf-1} \sum_{k=1}^{Nf-1} C_{\mu0}(\eta)  C_{\mu k}(\eta) x_k = \sum_{k=1}^{Nf-1} \delta_{0k}   x_k =0.
\end{equation}
Hence differentiating we obtain
\begin{equation}
    \sum_{\mu=0}^{Nf-1} \sum_{k=1}^{Nf-1} C_{\mu0}(\eta)  \frac{\partial C_{\mu k}(\eta)}{\partial \eta} x_k = - \sum_{\mu=0}^{Nf-1} \sum_{k=1}^{Nf-1} \frac{\partial C_{\mu 0}(\eta)}{\partial \eta}   C_{\mu k}(\eta) x_k.
\end{equation}
Finally differentiating Eq.~\eqref{Zero_mode_eigenvector} we have that
\begin{equation}
    \frac{\partial C_{\mu 0}(\eta)}{\partial \eta} = \frac{\mathrm{d}^2 \tilde{q}_\mu}{\mathrm{d} \eta^2}\Bigg/\left|\frac{\mathrm{d} \tilde{\mathbf{q}}}{\mathrm{d} \eta}\right| ,
\end{equation}
where we have used the fact that $\left|\frac{\mathrm{d} \tilde{\mathbf{q}}}{\mathrm{d} \eta}\right|$ is independent of $\eta$. The resulting expression for the Jacobian is then
\begin{equation}
    J =  \left|\frac{\mathrm{d} \tilde{\mathbf{q}}}{\mathrm{d} \eta}\right|-\sum_{\mu=0}^{Nf-1} \sum_{k=1}^{Nf-1} \left(\frac{\mathrm{d}^2 \tilde{q}_\mu}{\mathrm{d} \eta^2}\Bigg/\left|\frac{\mathrm{d} \tilde{\mathbf{q}}}{\mathrm{d} \eta}\right| \right)  C_{\mu k}(\eta) x_k,
\end{equation}
equivalent to Eq.~\eqref{Jacobian_Definition} in the main text.

\bibliography{references_extra,new} % please don't edit references.bib - if you want to add refs, add your own bib file

\end{document}

\title{Supporting information: Perturbatively corrected ring-polymer instanton theory for accurate tunneling splittings}
\author{Joseph E.\ Lawrence,$^*$ Jind\v{r}ich Du\v{s}ek and Jeremy O.\ Richardson}
\email{joseph.lawrence@phys.chem.ethz.ch \\  jeremy.richardson@phys.chem.ethz.ch}
\affiliation{Laboratory of Physical Chemistry, ETH Zurich, 8093 Zurich, Switzerland}
\date{\today}% It is always \today, today,

\maketitle

% \clearpage
\onecolumngrid

% \pagenumbering{arabic}% resets `page` counter to 1
\renewcommand{\thepage}{S\arabic{page}}
\renewcommand{\theequation}{S\arabic{equation}}
\renewcommand{\thefigure}{S\arabic{figure}}
\renewcommand{\thetable}{S\arabic{table}}
\renewcommand{\thesection}{S\arabic{section}}
\renewcommand{\thesubsection}{S\arabic{section}.\arabic{subsection}}
% % \renewcommand{\thepage}{S\arabic{page}}
% \setcounter{page}{1}
% \setcounter{figure}{0}
% \setcounter{equation}{0}

% \title{Explaining the efficiency of photosynthesis: 
% quantum uncertainty or classical vibrations? \\ Supplementary Materials}
% \author{Johan E.\ Runeson, Joseph E.\ Lawrence, Jonathan R.\ Mannouch, and Jeremy O.\ Richardson}

% \onecolumngrid

% \begin{center}
%     {\large Explaining the Efficiency of Photosynthesis: 
% Quantum Uncertainty or Classical Vibrations? \\ Supporting Information} \\[1em]
%     Johan E.\ Runeson, Joseph E.\ Lawrence, Jonathan R.\ Mannouch, and Jeremy O.\ Richardson^* \\
%     \emph{Laboratory of Physical Chemistry, ETH Zurich, 8093 Zürich, Switzerland}
% \end{center}
% \settitle
% \authors
% \thispagestyle{empty}
% \addtocounter{page}{-1}

% \maketitle

% \section*{Supplementary Materials}

\newcommand{\tot}{{\tau}}
\newcommand{\totzero}{{\tau_0}}
In the following we give tables presenting the convergence of the numerical results for malonaldehyde and its mono-deuterated isotopologue considered in the main paper. For both systems there are two tables, one containing the uncorrected RPI result, $\Delta_{\text{RPI}}(\hbar)$, and the other containing the correction factor, $\Delta_{\text{RPI+PC}}(\hbar)/\Delta_{\text{RPI}}(\hbar)$. Tables \ref{H_leading} and \ref{H_factor} correspond to the unsubstituted malonaldehyde, and tables \ref{D_leading} and \ref{D_factor} to the mono-deueterated isotopologue. 
%Let us point out the multiplicative structure of the perturbative correction in Eq.~\ref{Final_PCIT}. One can compute leading-order tunnelling splitting $\Delta_{\text{RPI}}$ and the multiplicative correction factor $\left(1+\hbar\left(\Gamma^{(-)}_1-\Gamma^{(+)}_1\right)\right) \equiv f$ separately and each of them may have different convergence properties.
%TODO use the same notation as paper

In each table, we extrapolate the results for a fixed $\tau$ to the $N\to\infty$ limit.
This was done by using a linear extrapolation of the last two data points plotted against $1/N^2$. That the results should have a $1/N^2$ convergence is expected based on the error of the Trotter splitting
used in the discretisation of the path integral, and we find this extrapolation scheme to be numerically robust. For the leading order term, $\Delta_{\text{RPI}}(\hbar)$, we also performed a simple extrapolation to the $\tau\to\infty$ limit. This was done by plotting the $N\rightarrow\infty$ results against $1/\tau$ and extrapolating using a linear fit to the final two points or a quadratic fit to all three, both of which gave the same result to the quoted level of precision. 

%We thus picked the lowest amount of datapoints such that their
%dependence on $1/N^2$ was linear and then extrapolated to $1/N^2 = 0$.
%Secondly, we observed the dependence of the extrapolated tunnelling splittings resp. correction factors on $1/\beta$. We then examined the resulting numbers and where possible extrapolated this dependence to $1/\beta=0$. We then multiplied the two values together to obtain $\Delta$.

%We found the extrapolation more consistent for the leading order tunnelling splitting than for the multiplicative factors. In the case of $\Delta_{\text{RPI}}$ for deuterium, the $N\to\infty$ linear fit provided good results and in the case of $\Delta_{\text{RPI}}$ for hydrogen we have also done a quadratic extrapolation including more datapoints which were not linear in $1/N^2$ and found agreement with the linear extrapolation. For the $N,\tau\to\infty$ extrapolation, we have taken the average of a linear and quadratic extrapolation of the three $N\to\infty$ values we had obtained in the previous extrapolation.

%In the case of the multiplicative factors, the $N\to\infty$ extrapolation was done with linear and quadratic fits and the results were in agreement. The $N,\tau\to\infty$ extrapolation however encountered issues. Regarding $f$ for hydrogen, we have done a linear extrapolation of only the values for $\hbar/(k_\mathrm{B}\tau) > 50$~K and a quadratic extrapolation of all of the values. Then we took the average. Regarding $f$ for deuterium, we observed oscillatory convergence so no straightforwad extrapolation was possible.

Finally, to obtain the reported RPI+PC values we multiply the $N,\tau\to\infty$ RPI result, $\Delta_{\text{RPI}}$, by the perturbative correction factor. This gives $\Delta_{\text{RPI+PC}} = 22.1$~cm$^{-1}$ for hydrogen transfer and $\Delta_{\text{RPI+PC}} = 2.96$~cm$^{-1}$ for deuterium transfer. Note that one obtains the same result, to quoted precision, regardless of which of the three values of $\tau$ are used to calculate the correction factor, hence we find the correction factors to be sufficiently converged with respect to $\tau$ without the need for extrapolation.

\begin{table}[p]
    \centering
    \caption{A convergence table for unsubstituted malonaldehyde (hydrogen transfer).}
    \begin{subtable}[t]{.4\linewidth}
        \caption{The leading order tunneling splitting,
$\Delta_{\text{RPI}}$, given in cm$^{-1}$.}
        \vspace*{.2cm}
        \begin{tabular}{r|ccc}
        \toprule
        \multicolumn{1}{c}{} & \multicolumn{3}{c}{$T=\hbar / (k_\mathrm{B} \tau)$} \\
\cmidrule{2-4}
\multicolumn{1}{r}{$N$} & 50 K & 25 K & 12.5 K \\
\midrule
128  & {17.477} & & \\
256  & {18.793} & {17.462} &\\
512  & {19.167} & {18.774} & {17.454} \\
1024 & {19.264} & {19.147} & {18.754} \\
2048 & {19.289} & {19.242} & {19.136} \\
\midrule
$N\to\infty$ & 19.297 & 19.274 & 19.261 \\
\midrule
 %\multicolumn{4}{c}{$N,\tau \to \infty$ = 19.246} \\
 \multicolumn{4}{c}{$N,\tau \to \infty$ = 19.25} \\
\bottomrule
\end{tabular}
\label{H_leading}
    \end{subtable}%
    \begin{subtable}[t]{.4\linewidth}
        \caption{A convergence table for the perturbative correction factor, $\Delta_{\text{RPI+PC}}/\Delta_{\text{RPI}}$ (dimensionless).}
        \vspace*{.2cm}
        \begin{tabular}{r|ccc}
            \toprule
            \multicolumn{1}{c}{} & \multicolumn{3}{c}{$T=\hbar / (k_\mathrm{B} \tau)$} \\
            \cmidrule{2-4}
\multicolumn{1}{r}{$N$} & 50 K & 25 K & 12.5 K \\
\midrule
128  & {1.1533} & & \\
256  & {1.1493} & {1.1518} & \\
512  & {1.1485} & {1.1477} & {1.1521} \\
1024 & {1.1483} & {1.1468} & {1.1479} \\
\midrule
$N\to\infty$ & 1.1483	& 1.1468	& 1.1465 \\
% \midrule
%  \multicolumn{4}{c}{$N,\tau \to \infty$ = 1.1464} \\
\bottomrule
    \end{tabular}
    \label{H_factor}
    \end{subtable}
\end{table}

\begin{table}[p]
    \centering
    \caption{A convergence table for mono-deuterated malonaldehyde (deuterium transfer).}
    \begin{subtable}[t]{.4\linewidth}
        \caption{The leading-order tunneling splitting,
$\Delta_{\text{RPI}}$, given in cm$^{-1}$.
%The $N\to\infty$ value was obtained by linear extrapolation of values corresponding to
%one $\hbar / (k_\mathrm{B} \tau)$ plotted against $1/N^2$.
%The $N,\tau\to\infty$ value was obtained by averaging the linear and quadratic extrapolation of the $N\to\infty$
%values plotted against $1/\beta$.
}
\vspace*{.2cm}
\centering
\begin{tabular}{r|ccc}
\toprule
\multicolumn{1}{c}{} & \multicolumn{3}{c}{$T=\hbar / (k_\mathrm{B} \tau)$} \\
\cmidrule{2-4}
\multicolumn{1}{r}{$N$} & 50 K & 25 K & 12.5 K \\
\midrule
64   & {2.282} & & \\
128  & {2.563} & {2.279} & \\
256  & {2.651} & {2.558} & {2.277} \\
512  & {2.675} & {2.645} & {2.557} \\
1024 & {2.681} & {2.669} & {2.644} \\
2048 & {2.682} & {2.675} & {2.667} \\
\midrule
$N\to\infty$ & 2.683 & 2.676 & 2.675 \\
\midrule
 %\multicolumn{4}{c}{estimated $N,\tau\to\infty$ = 2.672} \\
 \multicolumn{4}{c}{ $N,\tau\to\infty$ = 2.67} \\
\bottomrule
\end{tabular}
\label{D_leading}
    \end{subtable}%
\begin{subtable}[t]{.4\linewidth}
\caption{The perturbative correction factor, $\Delta_{\text{RPI+PC}}/\Delta_{\text{RPI}}$ (dimensionless). 
%The results are in cm$^{-1}$.
%The $N\to\infty$ value was obtained by linear extrapolation of values corresponding to
%one $\hbar / (k_\mathrm{B} \tau)$ plotted against $1/N^2$.
}
\vspace*{.2cm}
\centering
\begin{tabular}{r|ccc}
\toprule
\multicolumn{1}{c}{} & \multicolumn{3}{c}{$T=\hbar / (k_\mathrm{B} \tau)$} \\
\cmidrule{2-4}
\multicolumn{1}{r}{$N$} & 50 K & 25 K & 12.5 K  \\
\midrule
64   & {1.1144} & & \\
128  & {1.1095} & {1.1129} &  \\
256  & {1.1090} & {1.1079} & {1.1132} \\
512  & {1.1089} & {1.1073} & {1.1083} \\
1024 & {1.1088} & {1.1073} & {1.1077} \\
\midrule
$N\to\infty$	& 1.1088	& 1.1072	& 1.1075 \\
%$N,\tau \to \infty$ & \multicolumn{3}{c}{1.107 $\pm${} 0.005} \\
\bottomrule
\end{tabular}
\label{D_factor}
\end{subtable}
\end{table}

\bibliographystyle{achemso}
\bibliography{references} %,references}